\documentclass[a4paper,fleqn]{cas-dc}
\usepackage[authoryear,longnamesfirst]{natbib}

\usepackage{subcaption}  
\usepackage{xcolor}

\newcommand{\revone}[1]{\textcolor{black}{#1}}
\newcommand{\revtwo}[1]{\textcolor{black}{#1}}

\begin{document}
\let\WriteBookmarks\relax
\def\floatpagepagefraction{1}
\def\textpagefraction{.001}

\shorttitle{Neural-Parameterized Cellular}
\shortauthors{M. Zhenirovskyy et~al.}

\title [mode = title] {Neural-Parameterized Cellular Automata for Wildfire Spread}

\author[1]{Maksym Zhenirovskyy}
\cormark[1]
\ead{mzhenirovskyy@fujitsu.com}

\author[1]{Ion Matei}
\author[1]{Rohit Vuppala}
\author[1]{Takuya Kurihana}
\author[1]{Hon Yung Wong}

\affiliation[1]{organization={Fujitsu Research of America},
	addressline={4655 Great America Pkwy},
	city={Santa Clara},
	postcode={95124},
	state={CA},
	country={USA}}

\cortext[1]{Corresponding author}

\begin{abstract}
Traditional wildfire models rely on rigid, low-dimensional parameters and static fuel maps, frequently underpredicting fire spread. To address this weakness, we introduce a hybrid deep-learning parameterized Probabilistic Cellular Automata (CA) framework implemented in JAX. Our approach employs a Multi-Scale Convolutional Neural Network to dynamically generate spatially varying parameters that govern fire-spread probability, wind alignment, and slope influence. This hybrid design captures complex, nonlinear environmental interactions while preserving the physical interpretability of the underlying three-state CA. The JAX implementation enables hardware acceleration and gradient-based parameter calibration. \revtwo{Evaluated on six large-scale wildfires in the western United States}, the model maintains $IoU > 0.6$ over 72-hour forecast horizons \revone{after a 10-day data assimilation window during which the model is fitted incrementally to observed perimeters; the resulting forecast is a conditional projection of fire growth under the suppression regime already encoded in those observations.}
\end{abstract}

\begin{keywords}
	Wildfire simulation \sep Differentiable Cellular Automata \sep JAX-based software
\end{keywords}

\maketitle

\section{Introduction and Motivation}

Records from the California Department of Forestry and Fire Protection (CAL FIRE) document a severe escalation in wildfire scale, with over 4.3 million acres burned in the 2020 season alone \cite{calfire_2020_incidents}. The resulting ecological damage and emergency response costs motivate the development of high-fidelity models, amenable for real-time usage. Such models are essential for optimizing evacuation protocols and guiding the optimal deployment of firefighting resources \cite{mathur2026spatiotemporal}. Machine learning has been increasingly applied across wildfire science, from fuel characterization to behavior prediction \cite{jain2020review}; our work extends this trend with a hybrid physics-ML architecture.

Historically, operational wildfire forecasting has relied on semi-empirical models such as the 1972 Rothermel equations \cite{rothermel1972mathematical}. While foundational, these models depend on static, calibrated parameters and incur prohibitive computational costs when evaluated over high-resolution grids \cite{xia2025pytorchfire,cakir2025jaxwildfire}. In operational emergency contexts, where data are sparse and rapidly evolving, lightweight simulations are preferred over computationally exhaustive analytical models \cite{weinhouse2025leveraging}. Cellular Automata (CA) offer a computationally efficient abstraction of fire dynamics \cite{cakir2025jaxwildfire}, discretizing the landscape into a two-dimensional grid and governing propagation through local transition rules. However, conventional CA implementations apply uniform spread probabilities across heterogeneous environments, limiting predictive accuracy in the presence of complex topography, variable wind fields, and diverse fuel types \cite{weinhouse2025leveraging,xia2025pytorchfire}. Rule-based extensions such as the wind-dependent long-range propagation of \cite{freire2019cellular} partially address this but remain manually calibrated and difficult to generalize.

\revone{
Operational fire-management practice has long been dominated by simulators built on these semi-empirical equations. FARSITE \cite{finney1998farsite} propagates the fire perimeter as a vector polygon using Huygens' principle and elliptical wavelets, with vertex-level spread rates supplied by Rothermel's surface-fire model together with crown-fire and spotting submodels. The FlamMap analysis system \cite{finney2006flammap} bundles FARSITE with the Minimum Travel Time algorithm \cite{finney2002mtt}, a Treatment Optimization module, and conditional burn-probability tools. These simulators constitute the operational reference class against which new spread models are ultimately positioned, but they share two structural limitations relevant to the present work: they are not differentiable, so calibration to observed fires must be carried out by manual tuning of fuel adjustment factors or by outer-loop data assimilation (e.g., ensemble Kalman filtering); and their parameterization remains low-dimensional, with no native mechanism to learn high-dimensional spatially varying fields from observed perimeters.}

\revone{In parallel, the data-driven wildfire-modeling community has explored learned representations of fire spread that move outside the CA paradigm. Recent examples include deep-learning forecasters trained on time-series remote-sensing inputs \cite{lahrichi2025improved,anastasiou2025wildfire}, denoising-diffusion surrogates for probabilistic spread prediction \cite{yu2025probabilistic}, physics-informed world models for fire dynamics \cite{zhou2025physfire}, and hierarchical graph neural ODE approaches for global wildfire activity \cite{xu2026advanced}. These methods typically abandon the explicit state-transition structure of cellular automata in favor of fully learned dynamics, gaining representational flexibility at the cost of physical interpretability. Our hybrid model is positioned at the intersection of these two traditions: it retains the three-state probabilistic CA substrate so that mass conservation and ignition dynamics remain physically transparent, while replacing the historically static parameter set with spatially varying fields produced by an end-to-end differentiable neural network.}

\begin{figure}
    \centering
    \includegraphics[width=0.9\columnwidth]{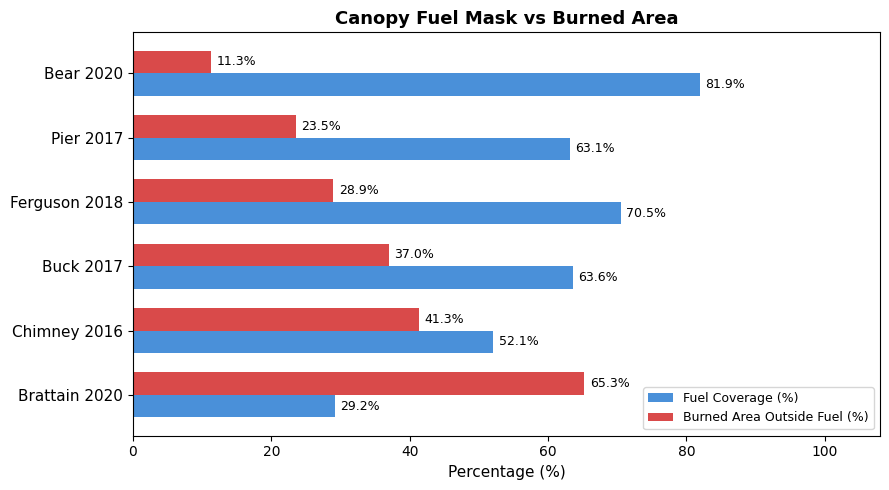}
    \caption{Canopy Fuel Mask vs  Burned Area. The charts show that the standard Canopy Cover (CC) and Canopy Bulk Density (CBD) do not capture the full extent of the burn.}
    \label{fig:canopy1}
\end{figure}

\begin{figure}
    \centering
    \includegraphics[width=0.9\columnwidth]{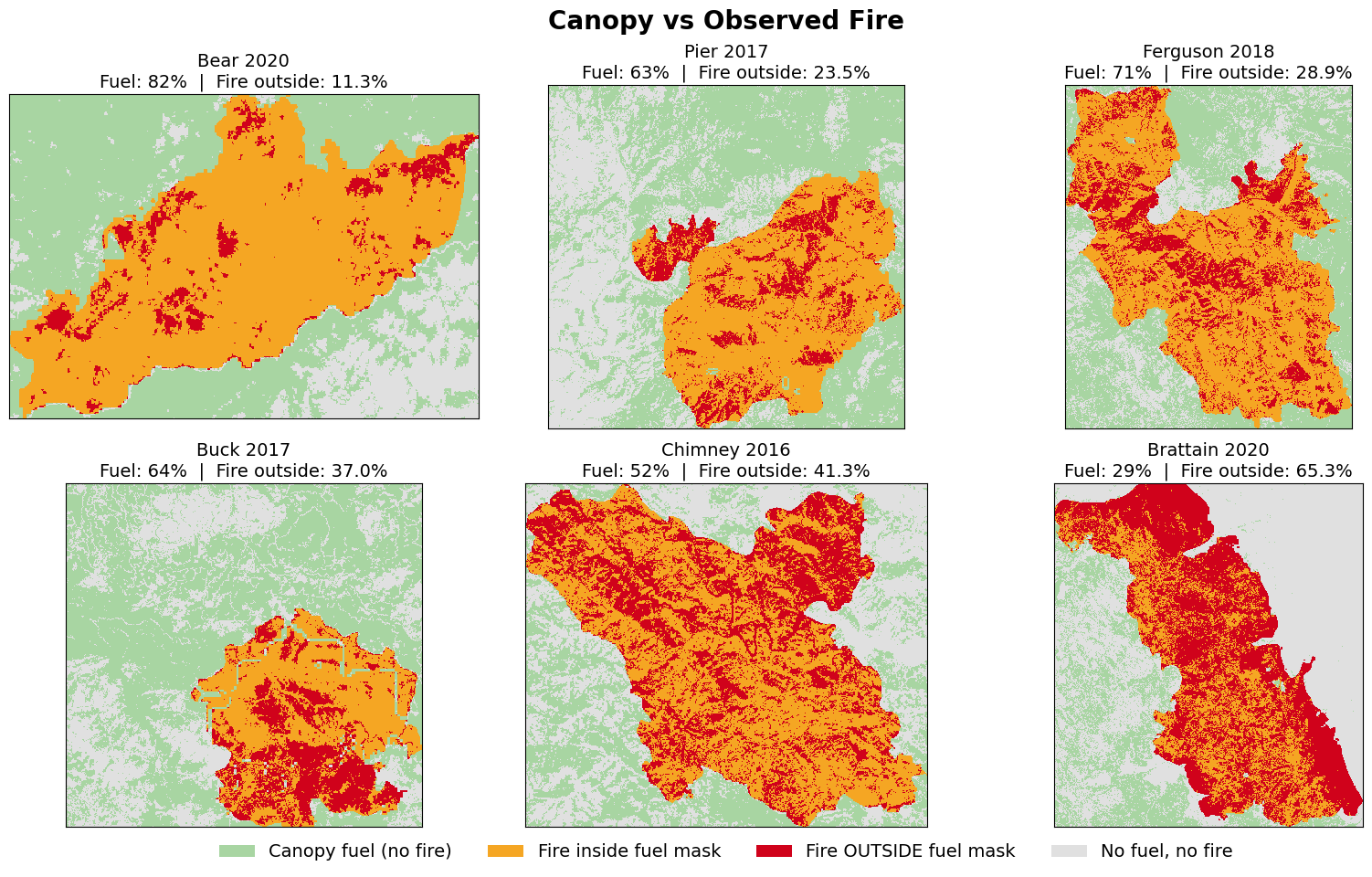}
    \caption{Spatial maps detailing the Canopy vs Observed Fire, highlighting the vast regions where fire propagated successfully without corresponding canopy fuel.}
    \label{fig:canopy2}
\end{figure}

A systemic limitation shared by both physics-based simulators and data-driven approaches is the reliance on static vegetation and canopy density metrics to define burnable boundaries \cite{chen2024autost}. Observational evidence indicates that models constrained to mapped fuel coverage consistently underpredict fire extent. Spatial analysis of recent events reveals that a substantial fraction of burned area falls in regions classified as lacking canopy or surface fuels (Figures \ref{fig:canopy1} - \ref{fig:canopy2}).

For instance, in the Chimney 2016 fire, 41.3\% of the final burned area lay outside mapped fuel coverage (Figures \ref{fig:canopy1} - \ref{fig:canopy2}). This pattern recurs across diverse biomes and fire events. For the Brattain Fire 2020, a staggering 65.3\% of the final fire footprint occurred in regions lacking categorized canopy vegetation, with only 29.2\% of the total event area formally covered by fuel. Similar failures of the Canopy Fuel Mask are observed in the Buck 2017 fire (37.0\% fire outside fuel), the Ferguson 2018 fire (28.9\% outside fuel), and the Pier 2017 fire (23.5\% outside fuel). Even in highly forested events such as the Bear 2020 fire, 11.3\% of the combustion occurred beyond the mapped vegetation boundaries. 

These discrepancies arise from physical processes not captured by fuel-contiguity assumptions, including thermal radiative preheating, firebrand lofting, ember spotting across non-combustible barriers, and ignition of sub-canopy duff layers. A predictive framework must therefore be capable of inferring propagation potential beyond mapped vegetation boundaries.

While \cite{zheng2017forest} replaced traditional CA transition rules with an extreme learning machine, their approach fails to capture convolutional spatial context and remains dependent on static fuel masks. \revone{To address these limitations, we propose a JAX-based \cite{jax2018github} hybrid model. We retain the three-state probabilistic CA substrate of \cite{cakir2025jaxwildfire} and \cite{xia2025pytorchfire}-including the Poisson-aggregated ignition probability and the mass-conserving state-transition equations-and depart from both works in three concrete respects, which together constitute our \textit{primary contribution}: (i) the per-event global scalar parameters used in prior differentiable CA work are replaced with dense, spatially varying fields ($p_{base}$, $\alpha_{w1}$, $\alpha_{w2}$, $\alpha_s$, $\gamma$, and $fuel\_factor$) generated by a Multi-Scale Convolutional Neural Network (MS-CNN) operating at three kernel scales over topography, fuel context, and meteorology; (ii) the canopy-derived burnability mask used in prior work is replaced by a learnable continuous fuel-embedding matrix that produces a unified $fuel\_factor$ per cell, enabling propagation into regions that canopy products classify as non-burnable; and (iii) a single set of network weights and embedding parameters is learned jointly across all six events, in contrast to \cite{xia2025pytorchfire}, which requires independent per-event calibration.} \revtwo{Evaluated on six historical wildfires in the western United States (five in California and the Brattain 2020 fire in Lake County, Oregon, which shares a direct border with northeastern California),} the system maintains an Intersection over Union $IoU > 0.6$ for all forecast days following a 10-day calibration period across five of the six events, accurately predicting burn patterns even in areas lacking canopy fuel. The remaining event (Buck 2017), whose perimeter was heavily shaped by confinement tactics and tactical firing operations, is discussed in Section~\ref{subsec:spatial_analysis}. \revone{The shared parameterization across events demonstrates that the MS-CNN learns generalizable environmental representations rather than memorizing individual fires.}

The rest of this paper is structured as follows. Section 2 details the mathematical framework, integrating the three-state Probabilistic CA with the CNN-generated parameter fields, and defines the core equations for fire spread and state transitions. Section 3 outlines the end-to-end differentiable model architecture and describes the multi-component loss function utilized during training. Section 4 describes the geospatial and meteorological input datasets and the variable-specific preprocessing pipeline that feeds both the CNN parameter generator and the CA physics engine. Section 5 presents a rigorous evaluation of the model against large-scale historical wildfires, analyzing both global performance metrics and detailed spatial divergences. Finally, Section 6 provides concluding remarks on the framework's efficacy and future applications.

\section{Mathematical Framework}

Our model builds upon the differentiable Probabilistic CA models of \cite{xia2025pytorchfire, cakir2025jaxwildfire} by integrating a physics engine with a neural parameter generator. The full pipeline, from environmental inputs through the CNN and the CA physics engine to the predicted fire probability field, is end-to-end differentiable, meaning that loss gradients with respect to all learnable parameters can be computed via automatic differentiation \cite{shen2023differentiable}.

Both \cite{xia2025pytorchfire} and \cite{cakir2025jaxwildfire} discretize the landscape into a two-dimensional grid in which each cell occupies one of three states: \textit{unburned}, \textit{burning}, or \textit{burned}. The models strictly enforce a probabilistic invariant at all times across every node in the spatial grid:
\begin{equation}
p_{unburned}(x,y,t)+p_{burning}(x,y,t)+p_{burned}(x,y,t)=1.    
\end{equation}

In \cite{xia2025pytorchfire}, following the foundational CA formulation of \cite{alexandridis2008cellular}, the probability of fire spreading from a given cell to its neighbors is calculated by multiplying a base probability by several scaling factors, such as vegetation type ($p_{veg}$), density ($p_{den}$), wind, and slope. \cite{cakir2025jaxwildfire} refines this approach by modeling directional propagation as an unbounded, dimensionless score known as potential, denoted by $\phi(u, \Delta)$, where $u$ denotes the burning source cell and $\Delta$ represents the directional offset vector to a neighboring cell, $\Delta = (\delta_x, \delta_y) \in \{-1, 0, 1\}^2$. This potential is based on static environmental data and incorporates modifiers for both canopy ignition ($\mu_{veg}$) and ground-fuel density ($\mu_{den}$):

\begin{align}
	\phi(u,\Delta) &= p_{base} \cdot (1+\mu_{veg}(u)) \cdot (1+\mu_{den}(u)) \nonumber\\
	&\quad \cdot\, \kappa_{wind}(u,\Delta) \cdot \kappa_{slope}(u,\Delta).
\end{align}

\cite{cakir2025jaxwildfire} interprets this propagation potential as the intensity rate of a Poisson process. It calculates the final ignition probability by aggregating the expected number of fire sparks arriving from a target cell's Moore neighbors.

A key limitation of these formulations is their dependence on static vegetation metrics (e.g., $p_{veg}$ or $\mu_{veg}$) to define burnable boundaries. To address the limitations of static vegetation metrics and homogeneous parameterization, our model replaces the global scalar parameters and hard-coded fuel masks with spatially varying fields dynamically generated by a MS-CNN. Our approach replaces these static mappings with a learnable fuel-embedding matrix that produces a unified  $fuel\_factor$ per cell. \revtwo{Because the CA contains no explicit terms for ember spotting, secondary ignitions, or suppression activity, $fuel\_factor$ is best understood as an \emph{effective} propagation modifier rather than a pure fuel property: it captures the contribution of the surface and canopy fuel itself together with sub-grid processes that the governing equations do not represent separately. We discuss the implications of this conflation in the Conclusions.}

This modification fundamentally improves the measure of contagion, given by the Propagation Potential ($\phi$), which we define as the directional heat energy transferred from source $u$ to target $x$:

\begin{align}
	\phi(u \to x) &= p_{base}(x) \cdot fuel\_factor(u) \nonumber\\
	&\quad \cdot \kappa_{wind}(u) \cdot \kappa_{slope}(u).
\label{eq:potential}
\end{align}

The Wind Factor ($\kappa_{wind}$) and Slope Factor ($\kappa_{slope}$) remain consistent with the physics-based formulations established in \cite{xia2025pytorchfire, cakir2025jaxwildfire}:

\begin{align}
	\kappa_{wind}(u) &= \exp\!\bigl(\alpha_{w1}(u) \cdot V(u)\bigr) \cdot \exp\!\bigl(\alpha_{w2}(u) \cdot V(u) \nonumber\\
	&\qquad \cdot [\cos(\theta_{spread}-\theta_{wind}(u))-1]\bigr), \\
	\kappa_{slope}(u) &= \exp\!\bigl(\alpha_s(u) \cdot \mathrm{slope}(u) \nonumber\\
	&\quad \cdot \cos(\theta_{spread} - \theta_{aspect}(u))\bigr).
	\label{eq:slope}
\end{align}

where

\begin{itemize}
	\item $p_{base}(x,y)$: Base fire spread probability.
	\item $\alpha_{w1}(x,y)$: Wind speed scaling.
	\item $\alpha_{w2}(x,y)$: Wind direction alignment.
	\item $\alpha_{s}(x,y)$: Slope influence.
	\item $slope(u)$: Terrain slope at the neighbor, normalized to [0, 1] as $degrees/90$.
	\item $\theta_{aspect}$: Upslope direction at the neighbor (LANDFIRE compass degrees converted to math radians).
	\item $\theta_{spread}$: Direction of propagation from the source cell $u$ to the target neighbor, in math-convention radians.
	\item \revtwo{$fuel\_factor(x,y)$: Learned effective fuel contribution.}
\end{itemize}

Once the individual potentials from all adjacent neighbors in the spatial neighborhood $N(x)$ are computed, they are aggregated into an absolute cumulative heat exposure value, denoted as $\lambda(x)$, which drives the Ignition Probability ($p_{ignite}$):

\begin{align}
\lambda(x)=\sum_{u \in N(x)}\phi(u \to x), \label{eq:lambda_x} \\
p_{ignite}(x)=1-\exp(-\gamma(x) \cdot \lambda(x)). \label{eq:p_ignite}
\end{align}

During each temporal micro-step, the transition matrices are computed simultaneously across the grid \cite{cakir2025jaxwildfire}:

\begin{align}
	newly\_burning(x)=p_{unburned}(x) \cdot p_{ignite}(x), \\
	burned\_now(x)=p_{burning}(x) \cdot (1-p_{continue}).
\end{align}

Here, $p_{continue} \in [0,1]$ denotes the probability that a burning cell remains in the burning state at the next micro-step; the complementary term $(1-p_{continue})$ is therefore the per-step rate of transition from burning to burned. Following \cite{xia2025pytorchfire, cakir2025jaxwildfire}, we treat $p_{continue}$ as a fixed scalar hyperparameter rather than a learned spatially varying field.

The master probability state tensors are subsequently updated via strict conservation of mass principles to prepare for the next micro-step \cite{cakir2025jaxwildfire}:

\begin{align}
	p_{unburned}^{t+1}(x) &= p_{unburned}^t(x) \nonumber\\
	&\quad -newly\_burning(x), \label{eq:p_t_plus_1_1}\\
	p_{burning}^{t+1}(x) &= p_{burning}^t(x)+newly\_burning(x) \nonumber\\
	&\quad -burned\_now(x), \label{eq:p_t_plus_1_2}\\
	p_{burned}^{t+1}(x) &= p_{burned}^t(x)+burned\_now(x). \label{eq:p_t_plus_1_3}
\end{align}

Ultimately, the observable probability of a cell containing active or extinguished fire is defined as the union of the active and terminal combustion states:

\begin{align}
P_{fire}(x,y,t) = p_{burning}(x,y,t) + p_{burned}(x,y,t).
\end{align}

\section{Proposed Hybrid Model}

\begin{figure*}
    \centering
    \includegraphics[width=0.9\textwidth]{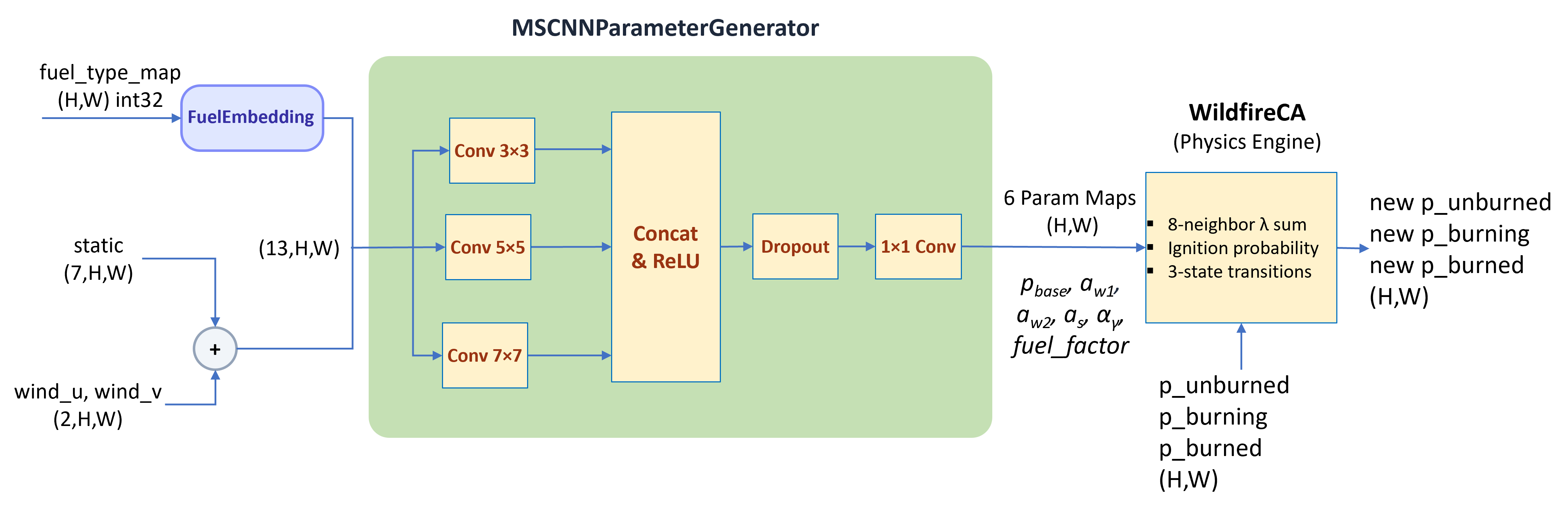}
    \caption{Detailed architectural diagram of the hybrid wildfire model combining a MS-CNN parameter
generator with a Probabilistic CA physics engine.}
    \label{fig:nn}
\end{figure*}

Our hybrid model, shown in Figure \ref{fig:nn}, is end-to-end differentiable and consists of the following main components:
\begin{itemize}
\item \textbf{FuelEmbedding}: Replaces categorical fuel rigidity with a learnable, continuous embedding matrix.
\item \textbf{MSCNNParameterGenerator}: Uses parallel convolutional pathways utilizing diverse kernel sizes (3x3, 5x5, and 7x7) to extract multi-scale topographical and meteorological features.
\item \textbf{WildfireCA}: A stateless physics engine that applies the parameter maps to execute the 8-neighbor potential summation.
\end{itemize}

To align predictions with daily satellite observations, the temporal execution is divided into 24-hour macro-steps, each containing multiple micro-steps. At the start of each macro-step, the \textbf{MSCNNParameterGenerator} ingests the current 13-channel input tensor. The tensor comprises the static LANDFIRE features, the learned fuel embeddings produced by \textbf{FuelEmbedding}, and the updated daily wind components. The module then outputs six spatially varying parameter maps: $p_{base}$, $\alpha_{w1}$, $\alpha_{w2}$, $\alpha_s$, $\gamma$, and $fuel\_factor$. These maps are then held fixed while the \textbf{WildfireCA} engine iterates through micro-steps within that macro-step. At each micro-step, the engine first computes the directional propagation potential $\phi(u \rightarrow x)$ from every Moore neighbor, as defined in Equation \ref{eq:potential}. It then aggregates the cumulative heat exposure $\lambda(x)$ using Equation \ref{eq:lambda_x} and derives the ignition probability $p_{ignite}(x)$ from Equation \ref{eq:p_ignite}. Finally, it updates the three-state probability tensor through Equations \ref{eq:p_t_plus_1_1}--\ref{eq:p_t_plus_1_3} to produce $\mathbf{p}^{t+1}$.

This pipeline differs from \cite{xia2025pytorchfire} and  \cite{cakir2025jaxwildfire} in two key respects. First, the parameter maps are not global scalars but dense spatial fields conditioned on multi-scale local context, allowing the model to express heterogeneous spread behavior across the domain. Second, the fuel contribution is not derived from static canopy masks but from the learned $fuel\_factor$, enabling propagation into regions that traditional formulations treat as non-burnable. Whereas \cite{xia2025pytorchfire, cakir2025jaxwildfire} use 15 micro-steps per macro-step, we found that 50 micro-steps produce the best performance, likely because the finer temporal resolution more closely approximates continuous propagation under spatially varying parameter fields.

\subsection{Loss Function}

For training, we use a composite objective enforcing pixel-level accuracy, area conservation, and geometric fidelity. We define the total loss for a single simulated day as a weighted sum of five complementary terms:

\begin{align}
	\mathcal{L}_{\mathrm{total}} &= w_{\mathrm{bce}}\,\mathcal{L}_{\mathrm{bce}} 
	+ w_{\mathrm{mse}}\,\mathcal{L}_{\mathrm{mse}} 
	+ w_{\mathrm{area}}\,\mathcal{L}_{\mathrm{area}} \nonumber\\
	&\quad + w_{\mathrm{iou}}\,\mathcal{L}_{\mathrm{iou}} 
	+ w_{\mathrm{pool}}\,\mathcal{L}_{\mathrm{pool}}.
	\label{eq:loss_total}
\end{align}

Each term addresses a distinct failure mode of the predicted fire probability field $\hat{P}(x,y) \in [0,1]$ relative to the observed fire $T(x,y) \in \{0,1\}$.

\paragraph{Frontier-Weighted Binary Cross-Entropy ($\mathcal{L}_{\mathrm{bce}}$):} In wildfire prediction, errors at the fire perimeter, where burned and unburned regions meet, are far more consequential for shape accuracy than errors in the interior or far field. Consequently, instead of applying the standard BCE for all locations, this loss focuses on the frontier region. We compute a frontier mask $F(x,y)$ as the set difference between the morphological dilation of the target fire $T(x,y)$ and the thresholded original fire region:
\begin{equation}
\label{eq:frontier_mask}
F = \mathcal{D}_{3\times3}(T) - \mathbf{1}[T > 0.5],
\end{equation}
where $\mathcal{D}_{3\times3}$ denotes the dilation operator and $\mathbf{1}[T > 0.5]$ represents the thresholding operation. The frontier-weighted BCE loss is then

\begin{align}
	\mathcal{L}_{\mathrm{bce}} &= \frac{1}{HW} \sum_{x,y} w(x,y) \nonumber\\
	&\quad \cdot \Bigl[ -T \log(\hat{P} + \varepsilon) \nonumber\\
	&\qquad - (1-T)\log(1 - \hat{P} + \varepsilon) \Bigr].
	\label{eq:bce}
\end{align}

where the weight map is defined as $w(x,y) = w_f$ for frontier pixels (i.e., for $F(x,y) > 0.5$), and $w(x,y) = 1$ elsewhere. The hyperparameter $w_f$ (set to 10 in our experiments) amplifies the gradient signal at the fire perimeter, steering the model toward accurate boundary delineation.

\paragraph{Mean Squared Error ($\mathcal{L}_{\mathrm{mse}}$):} The mean squared error (MSE) loss provides a smooth, pixel-wise regression signal:
\begin{equation}
\label{eq:mse}
\mathcal{L}_{\mathrm{mse}} = \frac{1}{HW} \sum_{x,y} \left(\hat{P}(x,y) - T(x,y)\right)^2,
\end{equation}
where $H$ and $W$ denote the height and width of the grid, respectively. MSE penalizes large deviations quadratically, stabilizing training when predictions diverge substantially from the target and complementing the logarithmic sensitivity of BCE.

\paragraph{Area Constraint ($\mathcal{L}_{\mathrm{area}}$):} The area constraint penalizes discrepancies in the total predicted burned area relative to the observed area:
\begin{equation}
\label{eq:area}
\mathcal{L}_{\mathrm{area}} = \frac{\left|\sum_{x,y} \hat{P}(x,y) - \sum_{x,y} T(x,y)\right|}{\sum_{x,y} T(x,y) + \varepsilon}.
\end{equation}
This term is scale-normalized by the target area, ensuring proportionate penalties regardless of fire size. Without it, the model may minimize pixel-wise loss while substantially misestimating total burned extent.

\paragraph{Soft IoU Loss ($\mathcal{L}_{\mathrm{iou}}$):} Pixel-level losses such as BCE and MSE decompose spatially and are insensitive to the global geometric configuration of the predicted fire region. A circular and an elongated prediction can yield comparable pixel-wise loss despite having different shapes. To provide explicit shape-matching gradients, we employ the soft Intersection-over-Union (Jaccard index) loss:


\begin{align}
	\mathrm{IoU} &= \frac{\mathcal{N}}{\mathcal{D} + \varepsilon}, 
	\label{eq:iou_calc}\\
	\mathcal{N} &= \sum_{x,y} \hat{P}(x,y) \cdot T(x,y), \nonumber\\
	\mathcal{D} &= \sum_{x,y} \hat{P}(x,y) + \sum_{x,y} T(x,y) 
	- \mathcal{N}. \nonumber
\end{align}

\begin{equation}
\label{eq:iou_loss}
\mathcal{L}_{\mathrm{iou}} = 1 - \mathrm{IoU}.
\end{equation}
The soft formulation operates directly on continuous probability maps, enabling gradient-based optimization without hard thresholding. The $IoU$ loss is weighted most heavily ($w_{\mathrm{iou}} = 4.0$) among all terms, reflecting the primacy of geometric accuracy in fire spread prediction. The numerical stability constant $\varepsilon = 10^{-7}$ is used in all logarithmic and division operations.

\paragraph{Pooled Mean Squared Error ($\mathcal{L}_{\mathrm{pool}}$):} The pooled MSE provides a coarse-grained spatial signal by applying average pooling to both the prediction and target fields prior to computing the squared error. Given a pooling window of size $k \times k$ with stride $k$ and valid (non-padded) boundaries:

\begin{equation}
\label{eq:pool}
\mathcal{L}_{\mathrm{pool}} = \mathrm{MSE}\!\left(\mathrm{AvgPool}_k(\hat{P}),\; \mathrm{AvgPool}_k(T)\right).
\end{equation}

The average pooling is implemented via convolution with a uniform $k \times k$ kernel normalized by $k^2$. This term smooths out fine-grained spatial noise and provides gradients that capture large-scale fire shape discrepancies, complementing the pixel-resolution MSE term.

\paragraph{Multi-Day Aggregation:} For multi-day simulations spanning $D$ days, each loss component is computed independently for every day and then averaged across the temporal dimension. Each event's per-day loss is further scaled by $1/D$ so that events of differing duration contribute equally to the epoch-level objective. The per-day losses are computed in parallel via vectorized mapping over the day index.

The default hyperparameter configuration used in our experiments is summarized in Table \ref{tab:loss_weights}.

\begin{table*}
	\centering
	\caption{Summary of Loss Weights}
	\label{tab:loss_weights}
\begin{tabular*}{\textwidth}{@{} LLLL@{} }
	\toprule
	\textbf{Term} & \textbf{Symbol} & \textbf{Weight} & \textbf{Role}\\
	\midrule
		Frontier-weighted $BCE$ & $\mathcal{L}_{\mathrm{bce}}$ & 0.5 & Pixel-level classification with perimeter emphasis \\
		Mean squared error & $\mathcal{L}_{\mathrm{mse}}$ & 0.1 & Smooth pixel-wise regression \\
		Area constraint & $\mathcal{L}_{\mathrm{area}}$ & 0.5 & Global burned-area conservation \\
		Soft $IoU$ & $\mathcal{L}_{\mathrm{iou}}$ & 4.0 & Geometric shape fidelity \\
		Pooled $MSE$ & $\mathcal{L}_{\mathrm{pool}}$ & 0.1 & Coarse-grained spatial matching \\
	\bottomrule
\end{tabular*}
\end{table*}

\section{Used Data and Preprocessing}

The model uses heterogeneous geospatial and meteorological inputs, with a unified tensor representation.

\subsection{Source Datasets}
In the following we list the used datasets:
\paragraph{Static data \cite{xia2025pytorchfire_data}:}
Seven static layers at 30\,m spatial resolution are drawn from the LANDFIRE
database.  These comprise Elevation
(\texttt{ELEV2020}, meters), Slope (\texttt{SLPD2020}, degrees), Aspect
(\texttt{ASP2020}, compass degrees), Canopy Bulk Density (\texttt{230CBD},
kg\,m$^{-3}\!\times 100$), Canopy Cover (\texttt{230CC}, percent), Canopy Height
(\texttt{230CH}, meters$\times 10$), and the Scott--Burgan 40-class Fire Behavior
Fuel Model (\texttt{230FBFM40}, categorical integers $0-202$).

\paragraph{Dynamic data:}
Dynamic meteorological forcing is provided by the ECMWF \mbox{ERA5-Land}
reanalysis~\cite{joaquin2019era5land}, specifically the $u$- and $v$-components
of the 10\,m wind field (m\,s$^{-1}$), stored as daily grids of shape
$(D, H, W)$ where $D$ is the number of simulated days.
\paragraph{Fire Perimeter Ground Truth:}
Daily observed fire perimeters are taken from the wildfire dataset published
by \cite{xia2025pytorchfire_data}, stored as a sequence of shape $(D+1, H, W)$ with values
in~$\{0,\dots,255\}$ rescaled to~$[0,1]$.

\subsection{Preprocessing and Normalization}

Each raw layer undergoes variable-specific transformations to ensure bounded,
physically meaningful inputs while preserving the information needed by both the CNN
and the CA.  No-data sentinels (values $\leq -9998$) are mapped to zero throughout. Details on relevant dataset variables are as follows:

\paragraph{Elevation:}
Per-event min--max normalization over valid cells maps elevation to $[0,1]$,
centering the local topographic range within each fire domain.
\paragraph{Slope:}
Two representations are derived from the raw degree values:
\begin{itemize}
  \item \emph{CNN channel:} degrees clipped to $[0,90]$, converted to radians,
        and passed through $\sin(\cdot)$ to yield a monotonically increasing,
        bounded feature.
  \item \emph{CA physics input:} degrees divided by~$90$ and clipped to $[0,1]$,
        preserving a linear encoding for the directional slope factor
        (Equation \ref{eq:slope}).
\end{itemize}
\paragraph{Aspect:}
LANDFIRE reports aspect in compass convention ($0^{\circ}\!=\!\text{North}$,
$90^{\circ}\!=\!\text{East}$, clockwise; $-1$ for flat cells).
\begin{itemize}
  \item \emph{CNN channels:} compass degrees are converted to radians and decomposed
        into $\sin$ and $\cos$ components, yielding two continuous channels that
        avoid the discontinuity at $0^{\circ}/360^{\circ}$.  Flat cells are set to
        zero in both channels.
  \item \emph{CA physics input:} the raw aspect is first flipped to the upslope
        direction by adding $180^{\circ}$, then converted from compass to
        mathematical convention ($0\!=\!\text{East}$, counter-clockwise) and
        expressed in radians on $[0,2\pi)$.
\end{itemize}

\paragraph{Canopy Variables:}
Canopy Bulk Density, Canopy Cover, and Canopy Height are divided by
domain-specific maxima ($45$, $100$, and $550$, respectively) to obtain
$[0,1]$-bounded features.  No-data cells are zeroed.

\paragraph{FBFM40 Fuel Model:}
Integer fuel-model codes are clipped to~$[0,202]$ and serve as indices into a
learnable embedding table.  A binary
\emph{fuel mask} $M_{\mathrm{fuel}}(x,y) \in \{0,1\}$ is derived by labeling codes
$91-99$ (urban, snow/ice, agriculture, water, bare ground) as non-burnable:
\begin{equation}
  M_{\mathrm{fuel}}(x,y) =
  \begin{cases}
    0, & \text{FBFM40}(x,y) \in [91,99],\\
    1, & \text{otherwise}.
  \end{cases}
  \label{eq:fuel_mask}
\end{equation}

This mask supplants the traditional canopy-derived burnability criterion
($\text{CC}>0 \wedge \text{CBD}>0$), which, as demonstrated in Figures~\ref{fig:canopy1} and~\ref{fig:canopy2},
systematically excludes grassland and shrubland cells where fire demonstrably
propagates.

\paragraph{Wind Components:}
The ERA5 $u$- and $v$-components are divided by a fixed maximum
$V_{\max}=10$\,m\,s$^{-1}$ for the CNN channels.  NaN values (missing data) are
replaced with zero, corresponding to calm conditions.  For the CA physics engine,
wind speed $V = \sqrt{u^2 + v^2}$ and direction $\theta_{\mathrm{wind}} =
\operatorname{atan2}(v,u) \bmod 2\pi$ are recomputed from the scaled components at
each daily macro-step.

\paragraph{CNN Input Tensor:} The Multi-Scale CNN receives a 13-channel tensor $\mathbf{X} \in
\mathbb{R}^{13 \times H \times W}$ constructed by concatenating three groups (Table~\ref{tab:cnn_channels}):

\begin{enumerate}
  \item \textbf{Seven static LANDFIRE channels} (indices $0-6$): normalized
        Elevation, $\sin(\text{slope})$, $\sin(\text{aspect})$,
        $\cos(\text{aspect})$, Canopy Bulk Density, Canopy Cover, and Canopy Height.
        These are computed once per event.
  \item \textbf{Four FBFM40 embedding channels} (indices $7-10$): produced at
        runtime by a learnable embedding table of size $203 \times 4$, initialized
        from $\mathcal{N}(0, 0.1)$.  Each cell's integer fuel-model code is mapped to
        a dense 4-dimensional vector, yielding a continuous spatial field of shape
        $(4, H, W)$.
        \label{sec:fuel_embed}
  \item \textbf{Two dynamic wind channels} (indices $11$--$12$): the normalized
        $u$- and $v$-components, updated at each 24-hour macro-step.
\end{enumerate}

\begin{table*}
\centering
\caption{CNN input channel specification.}
\label{tab:cnn_channels}
\begin{tabular*}{\textwidth}{@{} LLLL@{} }
\toprule
\textbf{Index} & \textbf{Variable} & \textbf{Source} & \textbf{Preprocessing} \\
\midrule
0  & Elevation         & ELEV2020  & Per-event min--max $\to [0,1]$ \\
1  & Slope             & SLPD2020  & $\sin(\text{clip}(\theta,0,90)\cdot\pi/180)$ \\
2  & Aspect (sin)      & ASP2020   & $\sin(\text{compass}\to\text{rad})$; flat$\to 0$ \\
3  & Aspect (cos)      & ASP2020   & $\cos(\text{compass}\to\text{rad})$; flat$\to 0$ \\
4  & Canopy Bulk Dens. & 230CBD    & $/\,45$ \\
5  & Canopy Cover      & 230CC     & $/\,100$ \\
6  & Canopy Height     & 230CH     & $/\,550$ \\
7--10 & Fuel Embedding & FBFM40    & Learnable $203\times 4$ table \\
11 & Wind $u$          & ERA5      & $/\,V_{\max}$; NaN$\to 0$ \\
12 & Wind $v$          & ERA5      & $/\,V_{\max}$; NaN$\to 0$ \\

\end{tabular*}
\end{table*}

\subsection{CA Physics Inputs}

The Probabilistic CA (Physics Engine, the leftmost part of Figure \ref{fig:nn}) operates on a separate set of physical
variables, preprocessed independently of the CNN channels to preserve their
native physical units within the governing equations \eqref{eq:potential}:

\begin{itemize}
  \item $\mathrm{slope}(u) \in [0,1]$: terrain slope normalized as
        $\mathrm{degrees}/90$.
  \item $\theta_{\mathrm{aspect}}(u) \in [0,2\pi)$: upslope direction in
        mathematical radians, converted from LANDFIRE compass convention.
  \item $V(u)$: wind speed (m\,s$^{-1}$), clipped to $[0,50]$.
  \item $\theta_{\mathrm{wind}}(u) \in [0,2\pi)$: wind direction computed as
        $\operatorname{atan2}(v,u) \bmod 2\pi$, following the mathematical
        convention ($0\!=\!\text{East}$, counter-clockwise).
  \item $M_{\mathrm{fuel}}(x,y)$: binary burnability mask defined
        in equation (\ref{eq:fuel_mask}), applied as a multiplicative gate on the
        ignition probability to enforce zero spread into non-burnable cells.
\end{itemize}

Boundary conditions are handled asymmetrically: fire intensity fields are
zero-padded (no fire enters from outside the domain), whereas environmental
fields-fuel factor, wind, slope, and aspect - are edge-padded to avoid
artificial boundary gradients.

\section{Results and Analysis}

We evaluate the model using a temporal split: the first 10 days of each event serve as a \revone{data-assimilation window during which the MS-CNN 	and the fuel embedding are fitted to observed perimeters as they arrive; 	thereafter, the parameters are frozen and the system generates a forward deterministic forecast for the subsequent days}. The historical calibration and forward deterministic forecasts were conducted using the wildfire perimeter dataset provided by \cite{xia2025pytorchfire_data}. \revone{The 10-day window is the midpoint of the 20-day analysis horizon and is an experimental design choice rather than an architectural constraint: the assimilation window can in principle begin at the first observed perimeter and grow incrementally as new perimeters arrive, so the operationally relevant lead time is set by perimeter observation latency rather than by a fixed 10-day requirement of the model. Suppression operations are ongoing throughout the assimilation phase, and because the 	observed perimeters used for fitting already encode the effects of firefighting (Section~\ref{subsec:spatial_analysis}), the forward forecast is a conditional projection of fire growth under the suppression regime that has taken hold in those observations, rather than a free-running simulation of an unsuppressed fire. This data-assimilation-then-forecast paradigm is shared with the differentiable-CA baselines 	\cite{xia2025pytorchfire,cakir2025jaxwildfire}, which similarly fit per-event parameters to observed perimeters before predictive use.}

\subsection{Global Performance Metrics}
\label{subsec:global_performance}

\begin{figure*}
	\centering
	\includegraphics[width=0.9\textwidth]{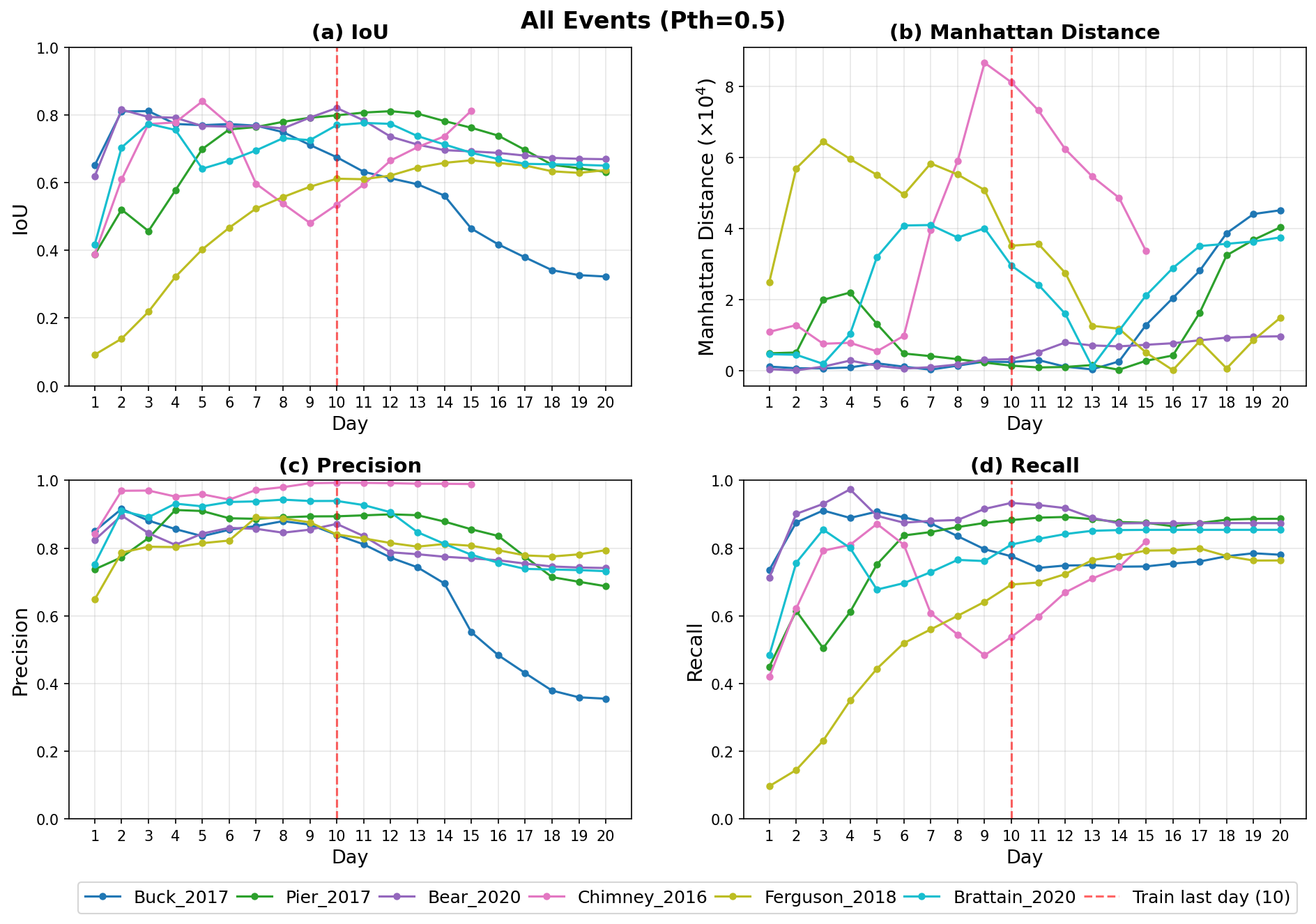}
	\caption{\revtwo{Per-event temporal trajectories of four spatial-agreement metrics over the 20-day analysis horizon for all six events: (a) Intersection over Union ($IoU$), (b) Manhattan Distance, (c) Precision, and (d) Recall. The vertical dashed line at Day~10 marks the end of the data-assimilation window; values to the right of this line correspond to forward forecasts produced with frozen parameters. All metrics are computed from binary burned/unburned masks obtained by thresholding the predicted probability map at $P_{\mathrm{th}}=0.5$. Precision (the fraction of predicted-burned cells that are observed burned) and Recall (the fraction of observed-burned cells that are predicted burned) decompose $IoU$ into complementary measures of over- and under-prediction, respectively, and quantify the false-positive (red) and false-negative (green) cells visualised in the per-event spatial-discrepancy figures of Section~\ref{subsec:spatial_analysis}.}}
	\label{fig:metrics}
\end{figure*}

As shown in Figure \ref{fig:metrics}, the architecture exhibits strong temporal stability. For most of the simulated events, including Bear, Chimney, Pier, and Brattain, the model consistently maintains an $IoU$ above 0.6 during the three critical days immediately after the 10-day calibration period. 

Our model outperforms \cite{xia2025pytorchfire}, which reported peak $IoU$ values of 0.408 (Bear 2020, day 8) and 0.525 (Pier 2017, day 12) on the two events evaluated in that study. By contrast, our model sustains $IoU > 0.6$ across the full 20-day simulation for both events.

\revtwo{Averaged across the full 20-day horizon, the model attains event-mean $IoU$ values of 0.74 (Bear~2020), 0.69 (Pier~2017), 0.69 (Brattain~2020), 0.66 (Chimney~2016), 0.61 (Buck~2017), and 0.52 (Ferguson~2018). Panels~(c) and~(d) of Figure~\ref{fig:metrics} decompose this agreement into Precision and Recall trajectories: five of the six events sustain both Precision and Recall above approximately 0.7 across the forecast horizon, with two notable deviations that are taken up in the per-event analysis of Section~\ref{subsec:spatial_analysis} -- a steep Precision decline for Buck~2017 after Day~10, and the expected low-Recall regime early in Ferguson~2018 caused by an unmodelled secondary ignition. Where the underlying error pattern is overprediction (red false-positive cells in the per-event spatial figures), Precision falls; where it is underprediction (green false-negative cells), Recall falls.}

\subsection{Detailed Spatial Analysis}
\label{subsec:spatial_analysis}

We analyze spatial error maps for the six events, classifying prediction errors as \textit{false positives} (overprediction) and \textit{false negatives} (underprediction). \revtwo{For each event we present three sub-panels covering Days 1--7, 8--14, and 15--20 of the simulation. Within every sub-panel the rows follow a fixed order: top, model prediction; middle, observed (target) perimeter from \cite{xia2025pytorchfire_data}; bottom, cell-wise overlap. In the overlap row, \textbf{red} cells are false positives (predicted burned, observed unburned - overprediction, where the simulated front runs ahead of the true perimeter) and \textbf{green} cells are false negatives (observed burned, predicted unburned - underprediction). These layout and color conventions apply to all six event figures (Figures~\ref{fig:bear_combined}--\ref{fig:buck_combined}).}

\begin{figure*}[htbp]
	\centering
	\includegraphics[width=0.9\textwidth]{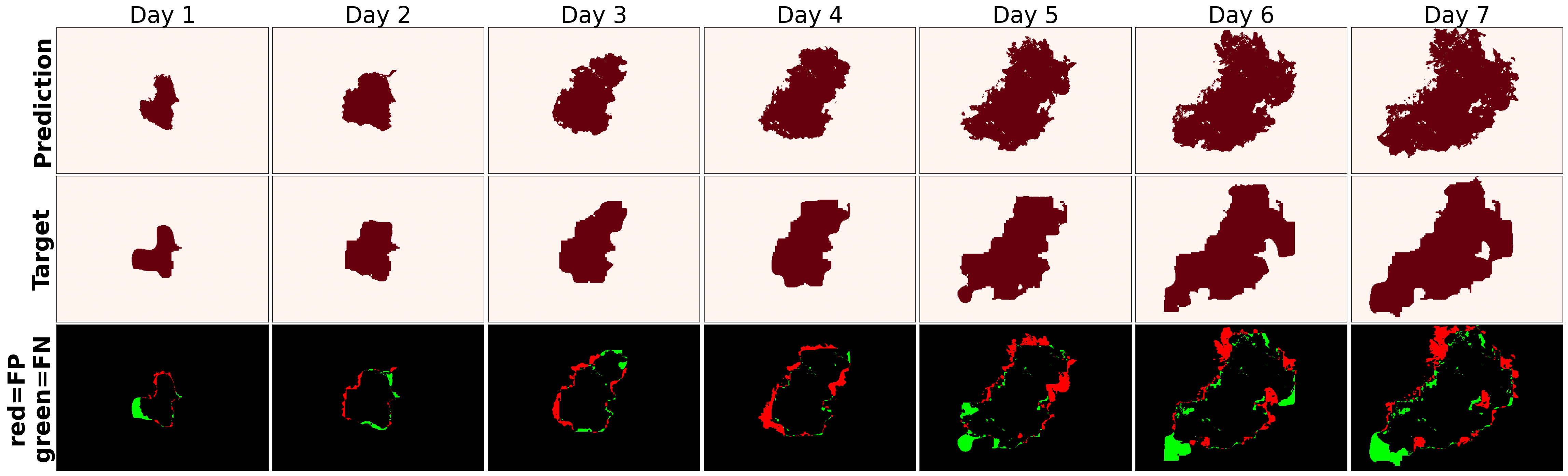}
	\caption{\revtwo{Bear 2020 fire analysis: (a) Days 1--7. Layout and color conventions as defined at the start of Section~\ref{subsec:spatial_analysis}.}}
	\label{fig:bear_combined}
\end{figure*}

\begin{figure*}[htbp]
	\ContinuedFloat
	\centering
	\includegraphics[width=0.9\textwidth]{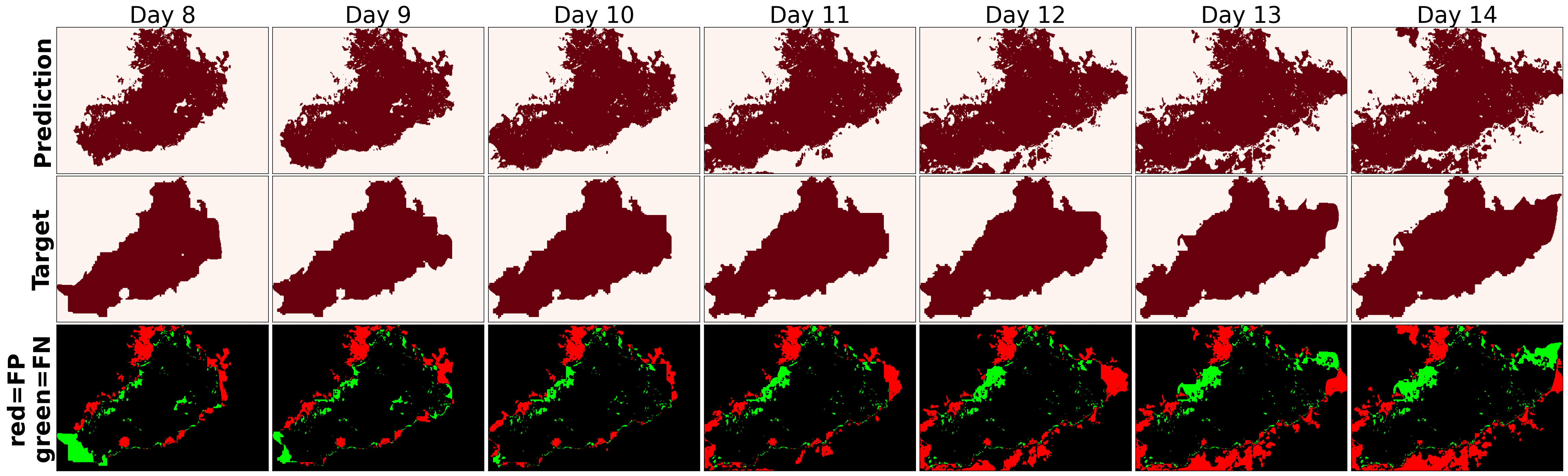}
	\caption{\revtwo{(cont.) Bear 2020. Days 8--14.}}
	\label{fig:bear_b}
\end{figure*}

\begin{figure*}[htbp]
	\ContinuedFloat
	\centering
	\includegraphics[width=0.9\textwidth]{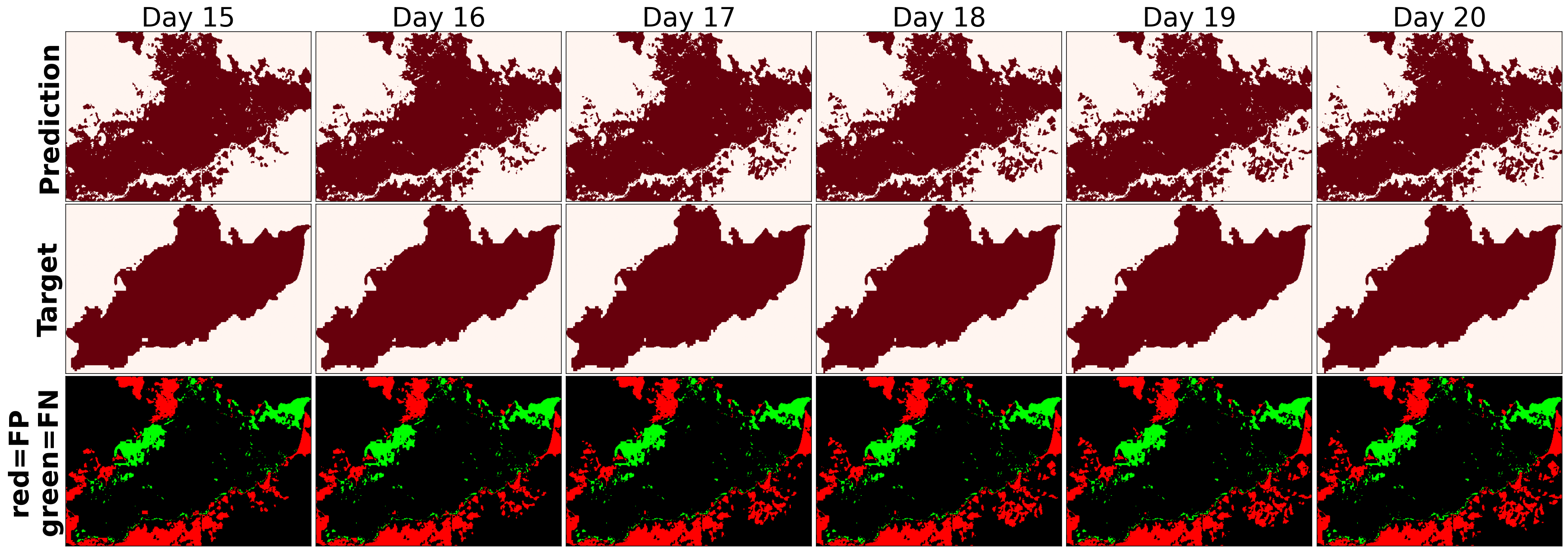}
	\caption{\revtwo{(cont.) Bear 2020. Days 15--20.}}
	\label{fig:bear_c}
\end{figure*}

\textbf{Bear 2020}: The model achieved strong agreement in aggregate burned-area metrics \revtwo{(event-mean $IoU = 0.74$, the highest among the six events; Figure~\ref{fig:metrics}(a))}. However, the spatial discrepancy plot (Figure \ref{fig:bear_combined}) reveals a concentration of False Positives along the eastern perimeter moving into Days 11-20, suggesting the MS-CNN over-indexed on the aggressive wind speed coefficient during the training phase\revtwo{; this is reflected in Figure~\ref{fig:metrics}(c) as a gradual Precision decline through the forecast horizon, while Recall remains stable above 0.85 (Figure~\ref{fig:metrics}(d))}.

\begin{figure*}[htbp]
	\centering
	\includegraphics[width=0.9\textwidth]{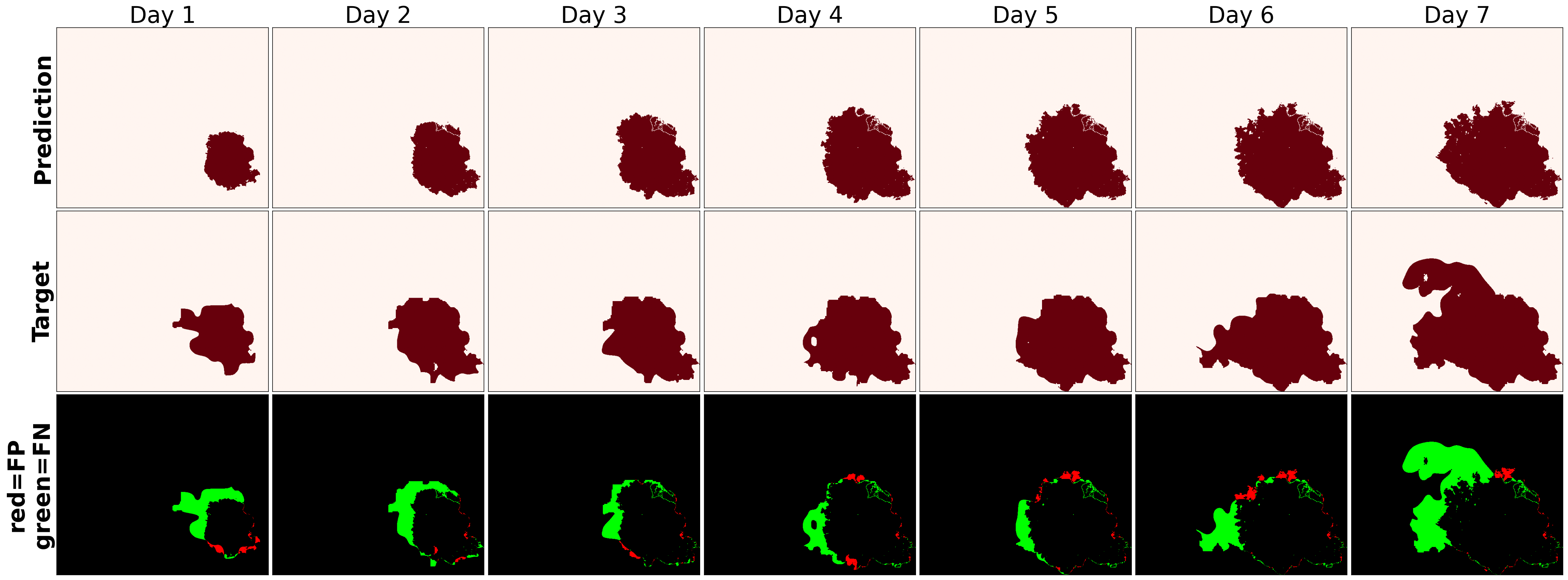}
	\caption{\revtwo{Chimney 2016 fire analysis: (a) Days 1--7. Layout and color conventions as defined at the start of Section~\ref{subsec:spatial_analysis}.}}
	\label{fig:chimney_combined}
\end{figure*}

\begin{figure*}[htbp]
	\ContinuedFloat
	\centering
	\includegraphics[width=0.9\textwidth]{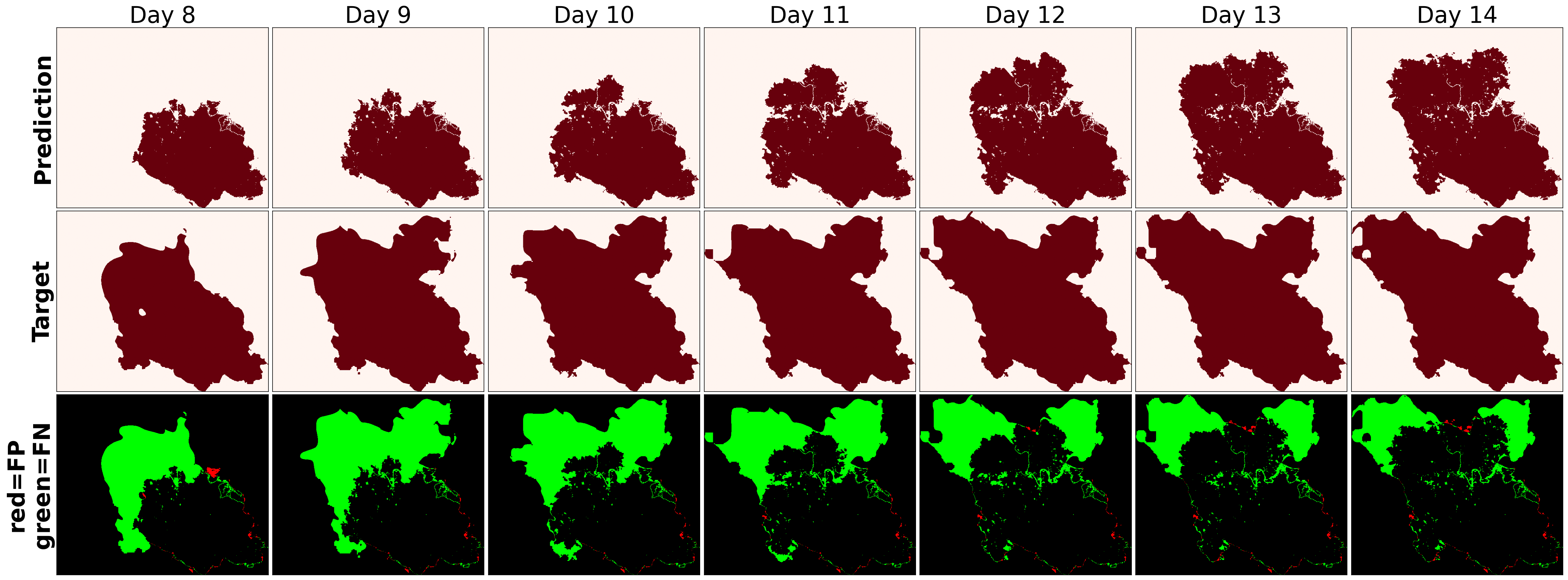}
	\caption{\revtwo{(cont.) Chimney 2016. Days 8--14.}}
	\label{fig:chimney_b}
\end{figure*}

\textbf{Chimney 2016}: This event demonstrates the advantage of learned fuel embeddings over static vegetation masks. Despite 41.3\% of the fire burning outside vegetative boundaries, the model successfully learned to assign non-zero fuel factors to these unburnable zones\revtwo{; consistent with this, Chimney sustains the highest Precision among all events in Figure~\ref{fig:metrics}(c), indicating that the learned fuel factors did not generate spurious overprediction in the unmapped fuel zones}. The discrepancy map (Figure \ref{fig:chimney_combined}) shows exceptional boundary adherence from Day 11 to 14\revtwo{, coinciding with the $IoU$ peak visible for this event in Figure~\ref{fig:metrics}(a) (event-mean $IoU = 0.66$)}.

\begin{figure*}[htbp]
	\centering
	\includegraphics[width=0.9\textwidth]{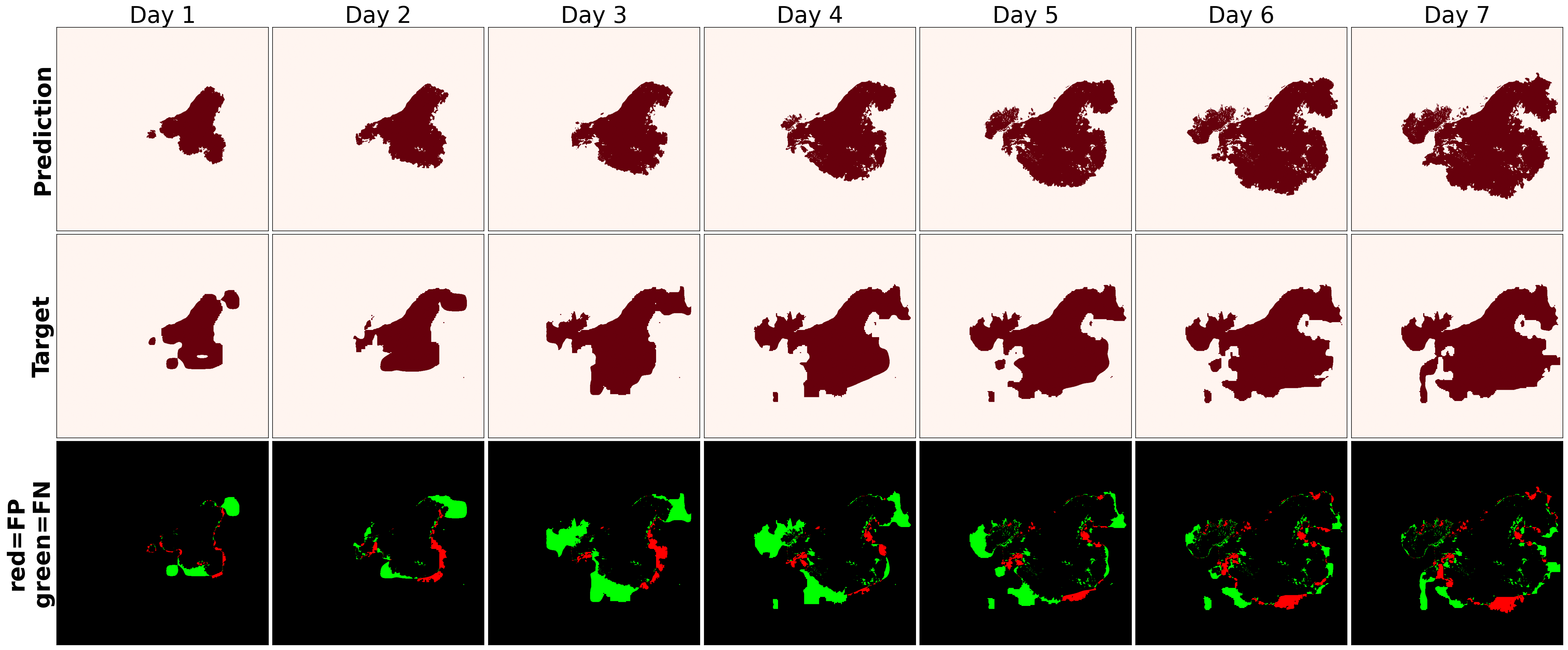}
	\caption{\revtwo{Pier 2017 fire analysis: (a) Days 1--7. Layout and color conventions as defined at the start of Section~\ref{subsec:spatial_analysis}.}}
	\label{fig:pier_combined}
\end{figure*}

\begin{figure*}[htbp]
	\ContinuedFloat
	\centering
	\includegraphics[width=0.9\textwidth]{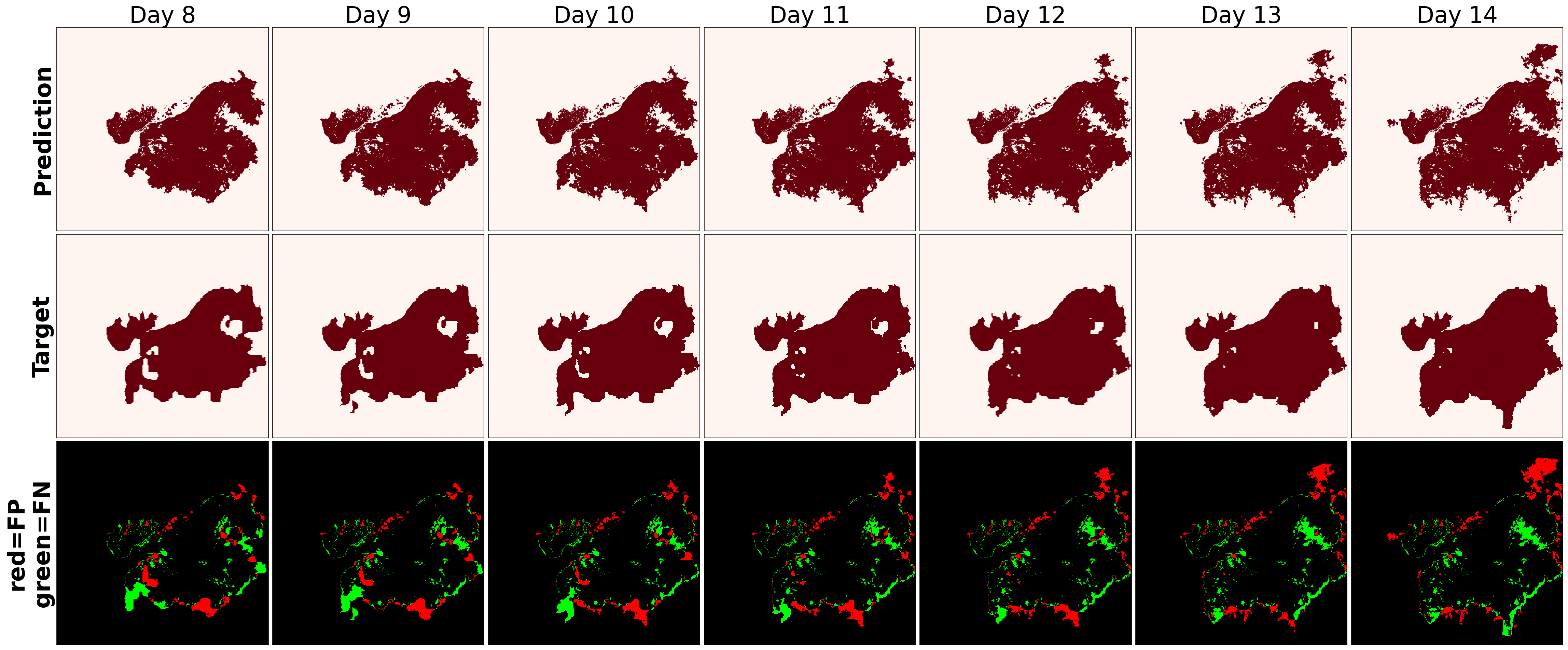}
	\caption{\revtwo{(cont.) Pier 2017. Days 8--14.}}
	\label{fig:pier_b}
\end{figure*}

\begin{figure*}[htbp]
	\ContinuedFloat
	\centering
	\includegraphics[width=0.9\textwidth]{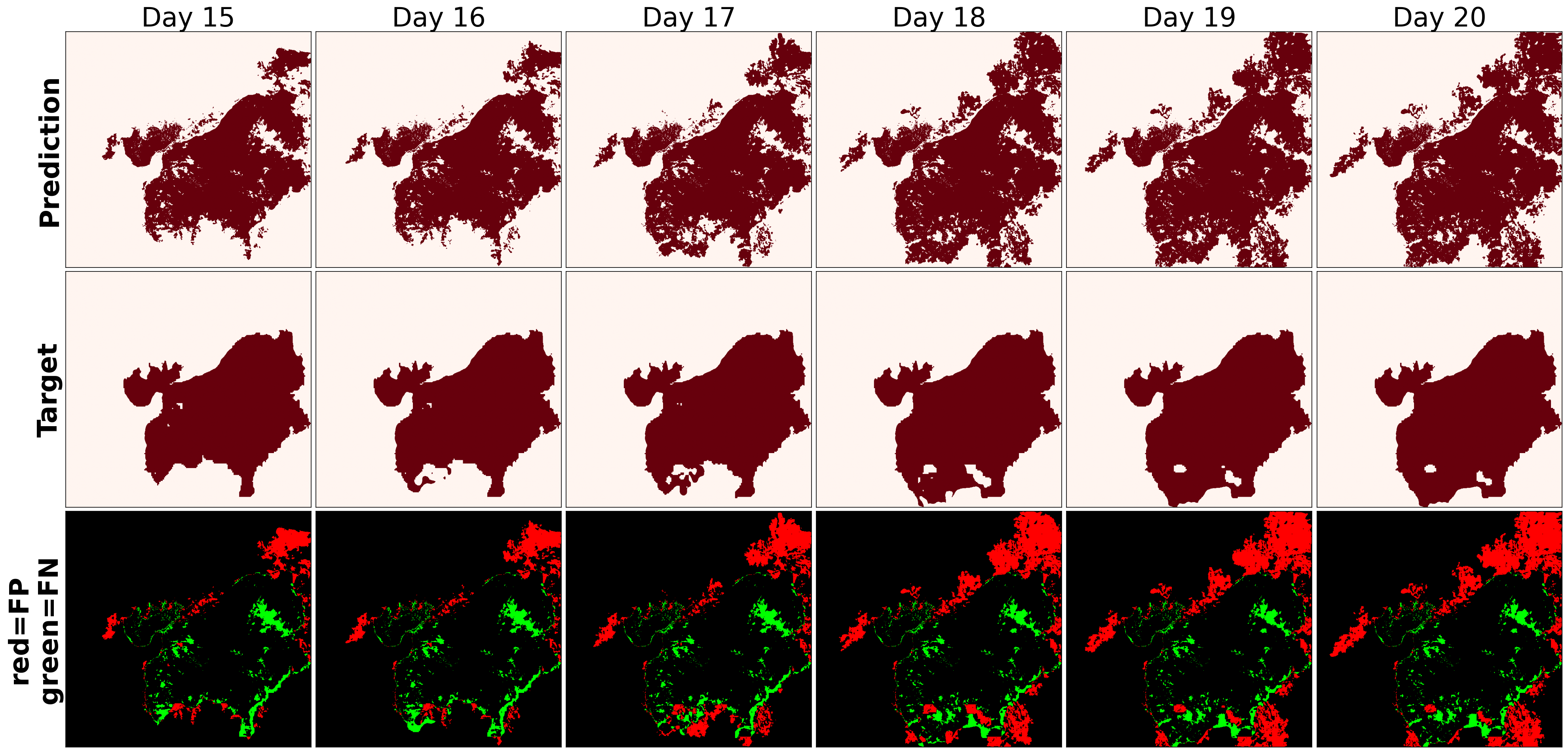}
	\caption{\revtwo{(cont.) Pier 2017. Days 15--20.}}
	\label{fig:pier_c}
\end{figure*}

\textbf{Pier 2017}: Characterized by extremely steep terrain, the target versus simulation overlay reveals high macro-structural agreement (Figure \ref{fig:pier_combined}) \revtwo{(event-mean $IoU = 0.69$, with both Precision and Recall remaining above 0.8 across the 20-day horizon in Figure~\ref{fig:metrics}(c,d))}. Localized green False Negatives on the southern flank suggest the model underpredicted the specific rate of downhill backing fire, a notoriously stochastic physical process \cite{xia2025pytorchfire}.

\begin{figure*}[htbp]
	\centering
	\includegraphics[width=0.9\textwidth]{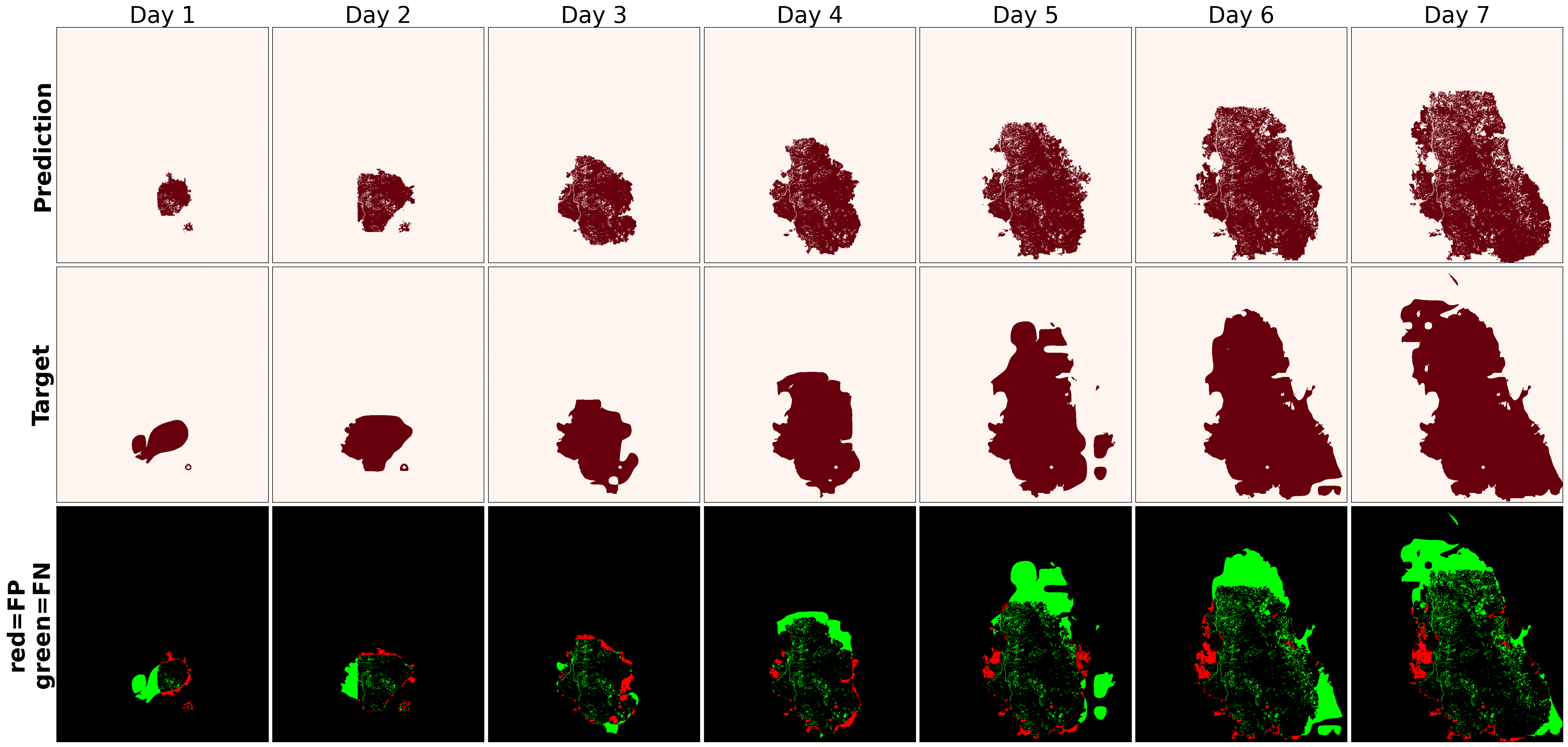}
	\caption{\revtwo{Brattain 2020 fire analysis: (a) Days 1--7. Layout and color conventions as defined at the start of Section~\ref{subsec:spatial_analysis}.}}
	\label{fig:brattain_combined}
\end{figure*}

\begin{figure*}[htbp]
	\ContinuedFloat
	\centering
	\includegraphics[width=0.9\textwidth]{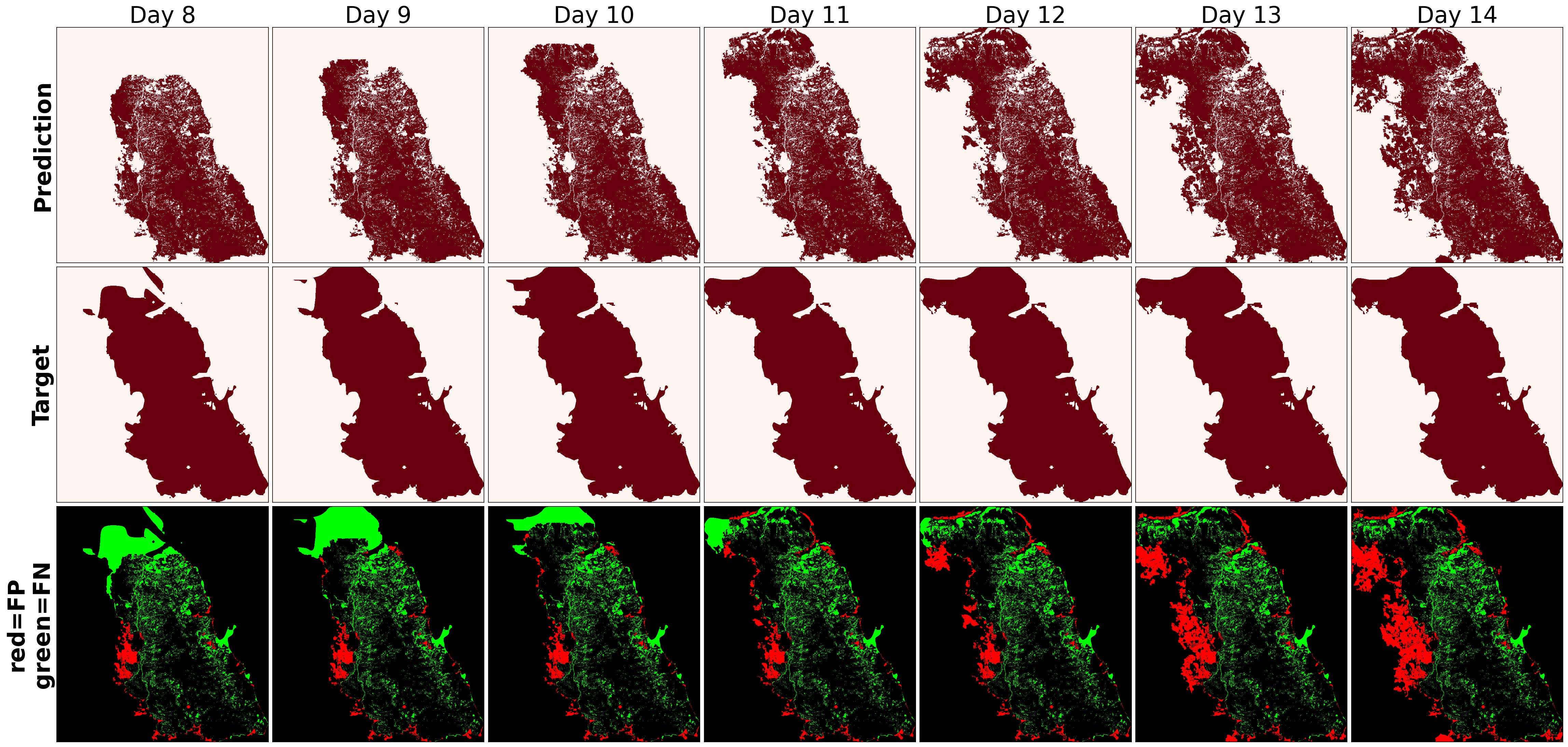}
	\caption{\revtwo{(cont.) Brattain 2020. Days 8--14.}}
	\label{fig:brattain_b}
\end{figure*}

\begin{figure*}[htbp]
	\ContinuedFloat
	\centering
	\includegraphics[width=0.9\textwidth]{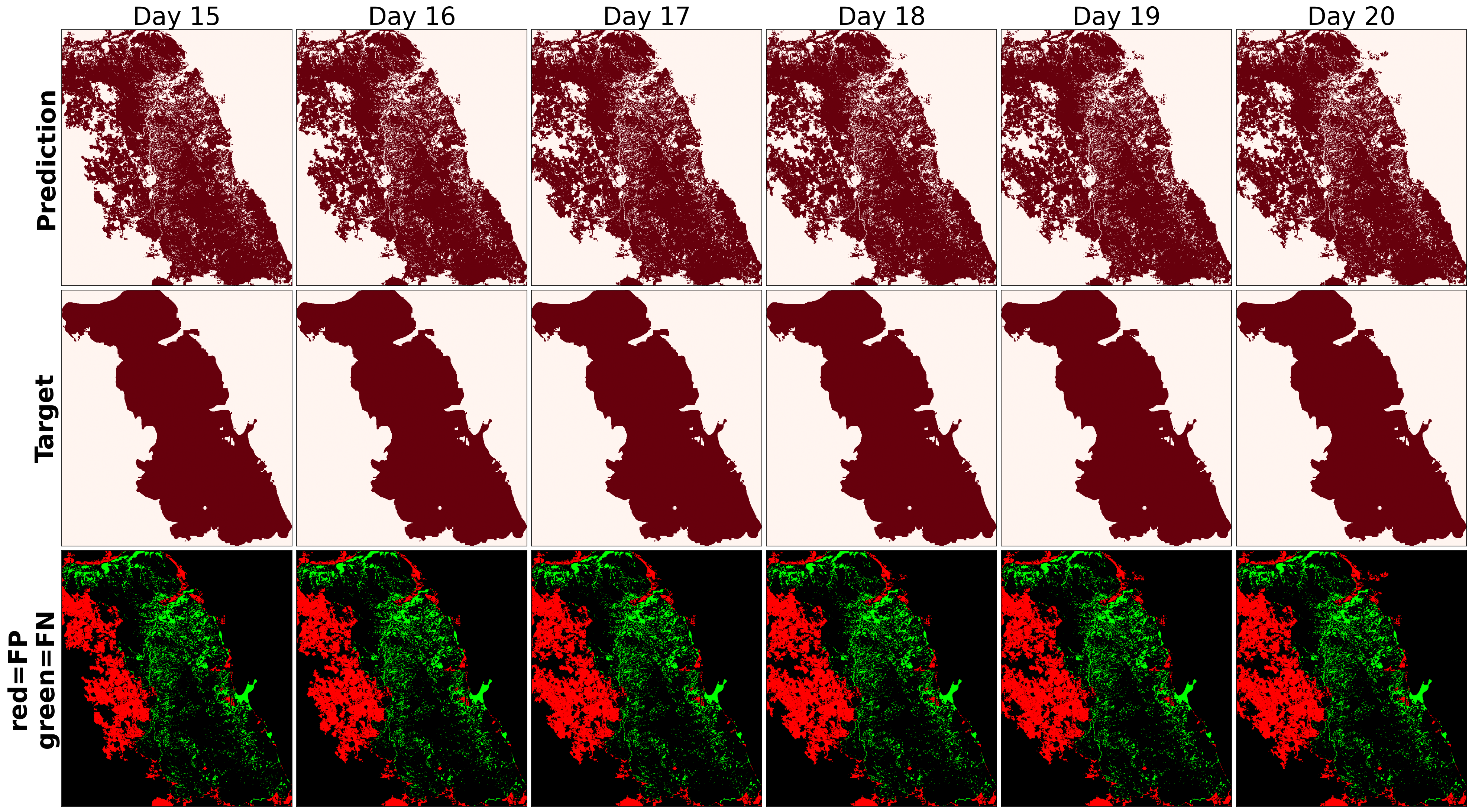}
	\caption{\revtwo{(cont.) Brattain 2020. Days 15--20.}}
	\label{fig:brattain_c}
\end{figure*}

\textbf{Brattain 2020}: \revtwo{This event-the only non-California fire in our evaluation set-burned in sagebrush--juniper with stringers of Ponderosa pine in Lake County, Oregon (south of Paisley, on the Paisley Ranger District of the Fremont--Winema National Forest), immediately north of the California state line, providing a partial test of cross-ecoregion behavior.} This event represents the most extreme test of the learned fuel embeddings, with 65.3\% of the final fire footprint occurring outside mapped canopy vegetation. Despite this, the model maintains strong geometric agreement throughout the simulation (Figure \ref{fig:brattain_combined}) \revtwo{(event-mean $IoU = 0.69$; Precision and Recall both remain above approximately 0.75 across the 20-day horizon, Figure~\ref{fig:metrics}(c,d))}.

\begin{figure*}[htbp]
	\centering
	\includegraphics[width=0.9\textwidth]{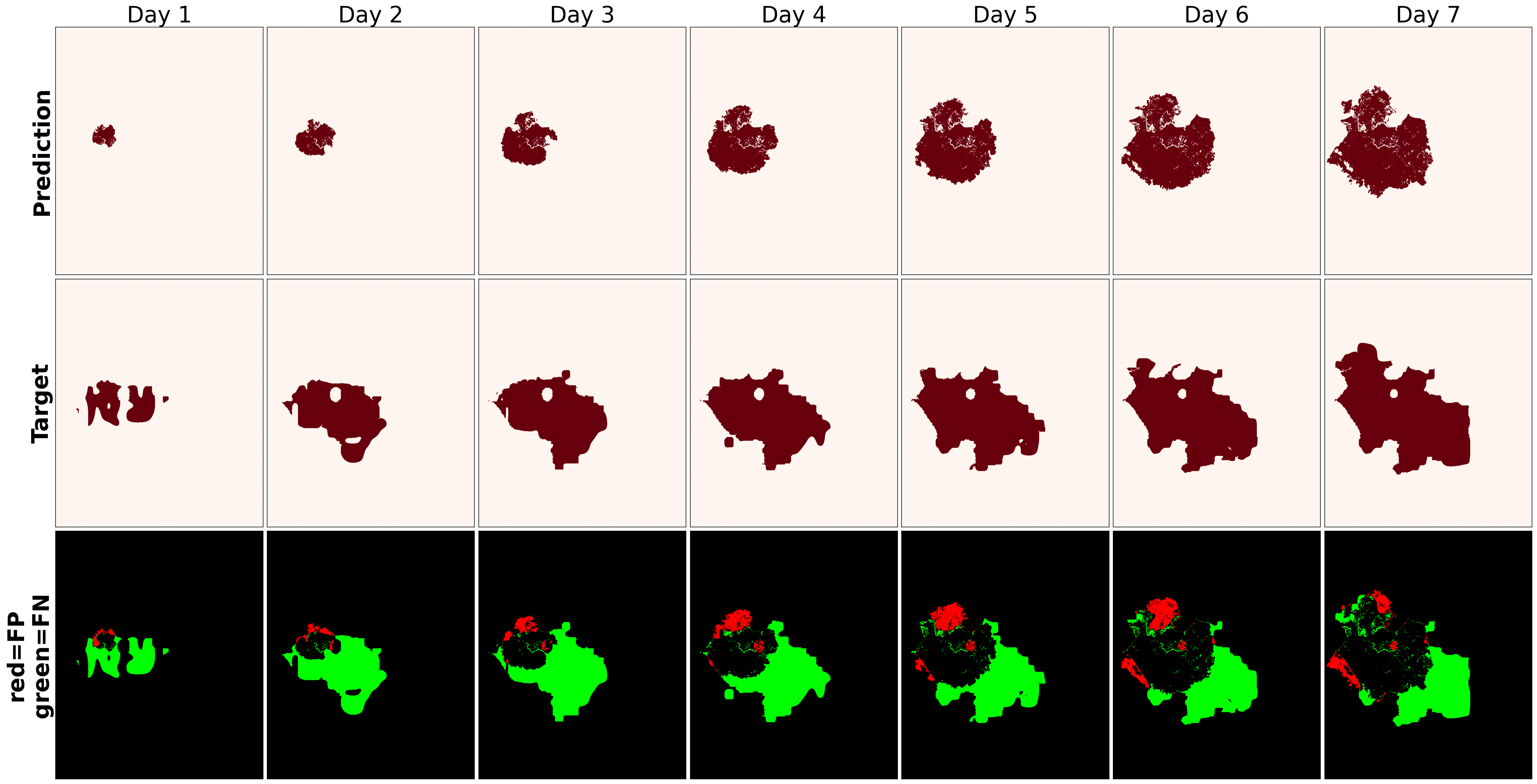}
	\caption{\revtwo{Ferguson 2018 fire analysis: (a) Days 1--7. Layout and color conventions as defined at the start of Section~\ref{subsec:spatial_analysis}.}}
	\label{fig:ferguson_combined}
\end{figure*}

\begin{figure*}[htbp]
	\ContinuedFloat
	\centering
	\includegraphics[width=0.9\textwidth]{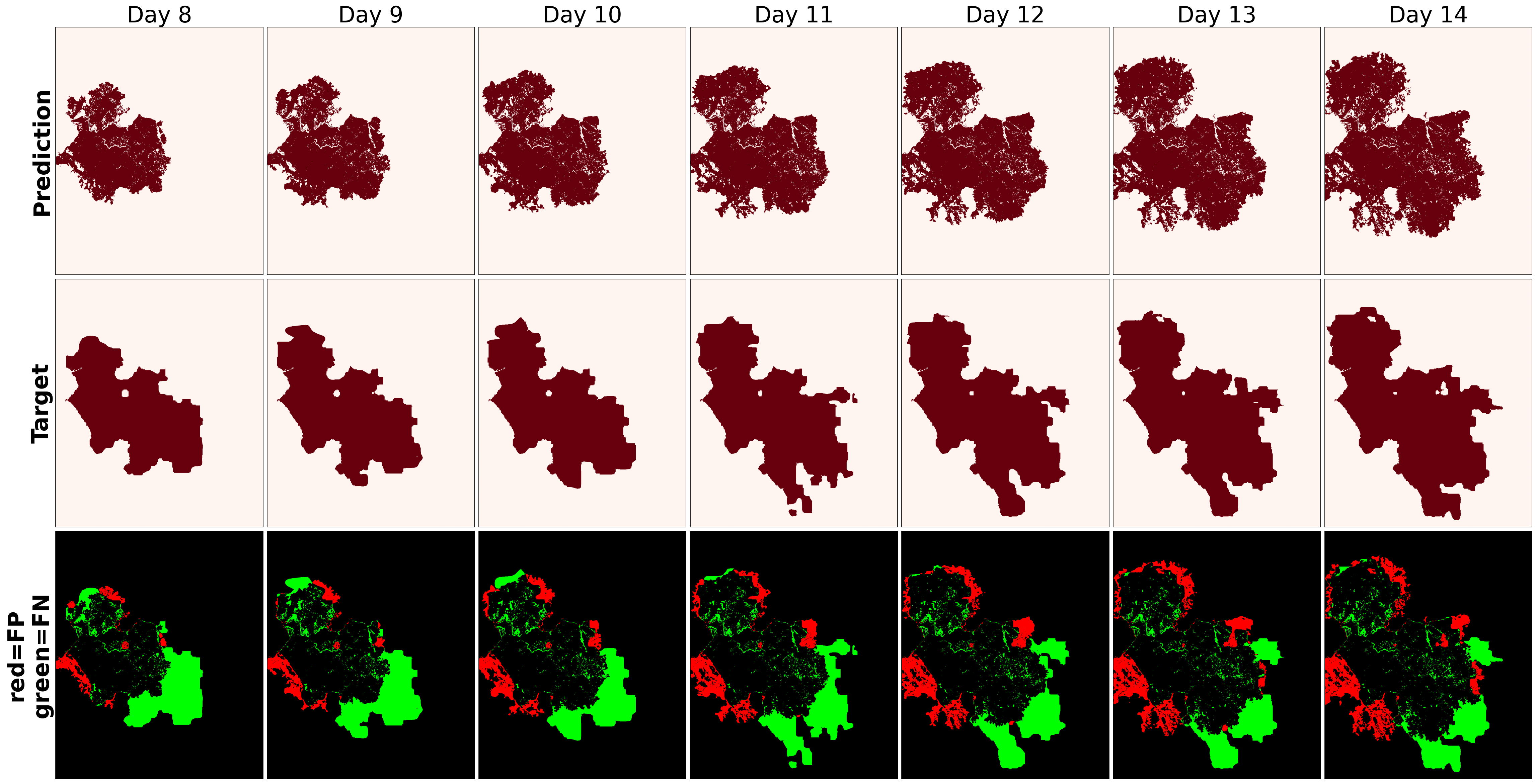}
	\caption{\revtwo{(cont.) Ferguson 2018. Days 8--14.}}
	\label{fig:ferguson_b}
\end{figure*}

\begin{figure*}[htbp]
	\ContinuedFloat
	\centering
	\includegraphics[width=0.9\textwidth]{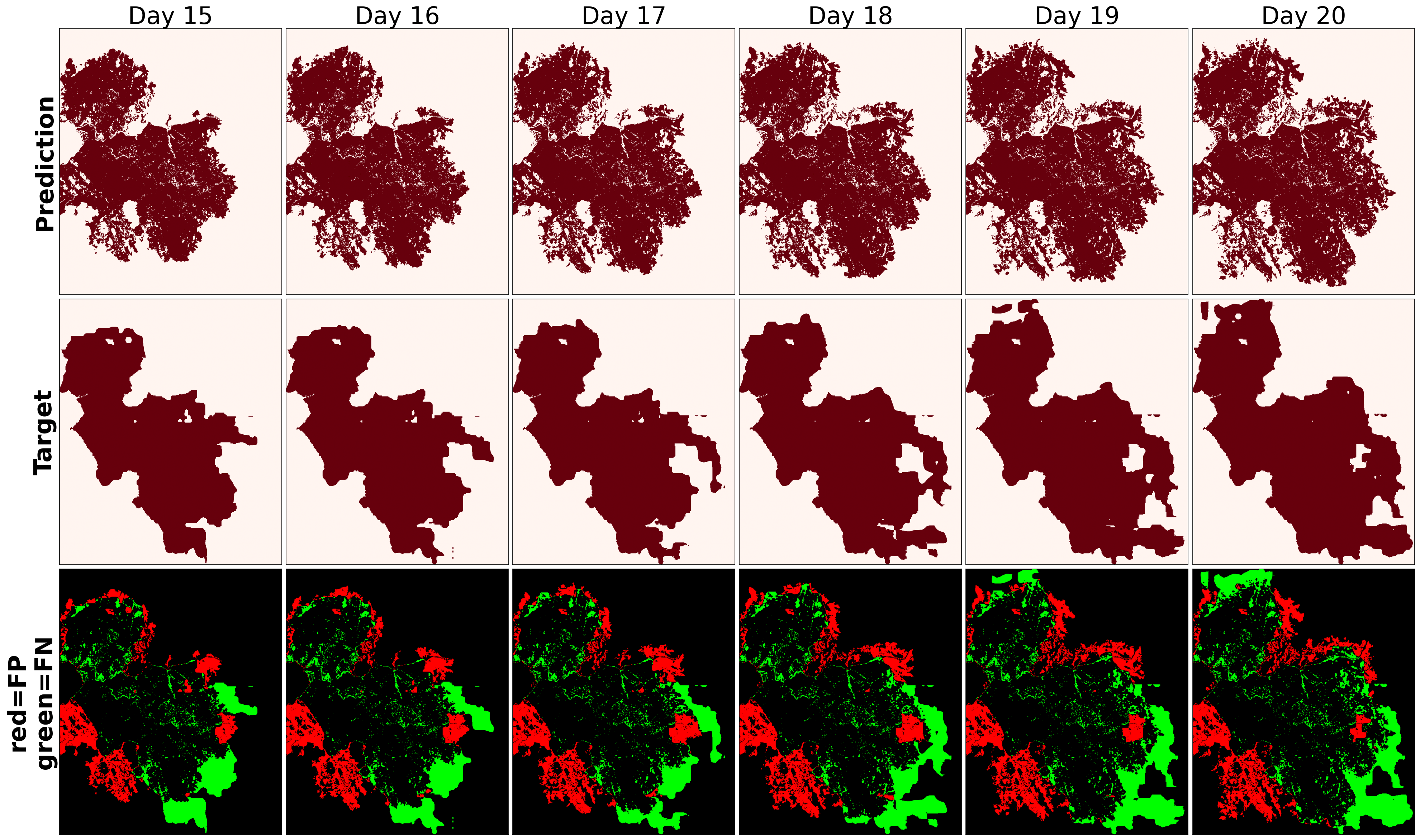}
	\caption{\revtwo{(cont.) Ferguson 2018. Days 15--20.}}
	\label{fig:ferguson_c}
\end{figure*}

\textbf{Ferguson 2018} is a notable outlier \revtwo{(event-mean $IoU = 0.52$, the lowest of the six events)}. The initial $IoU$ is exceptionally low (around 0.15 on Day 1, Figure \ref{fig:metrics}). This discrepancy was caused by a second, independent ignition event occurring geographically distinct from the primary fire front. Because the CA propagates fire only from active cells, it cannot model spontaneous secondary ignitions. \revtwo{The signature of this failure mode is unambiguous in Figure~\ref{fig:metrics}(d): Ferguson 2018 begins with Recall below 0.1 -- the model misses the entire secondary-ignition footprint -- and recovers steadily as the two fronts merge.} However, once the primary and secondary fires merged, the neural parameter generator rapidly recalibrates, and the $IoU$ \revtwo{recovers to above 0.6} by Day 10.

\begin{figure*}[htbp]
	\centering
	\includegraphics[width=0.9\textwidth]{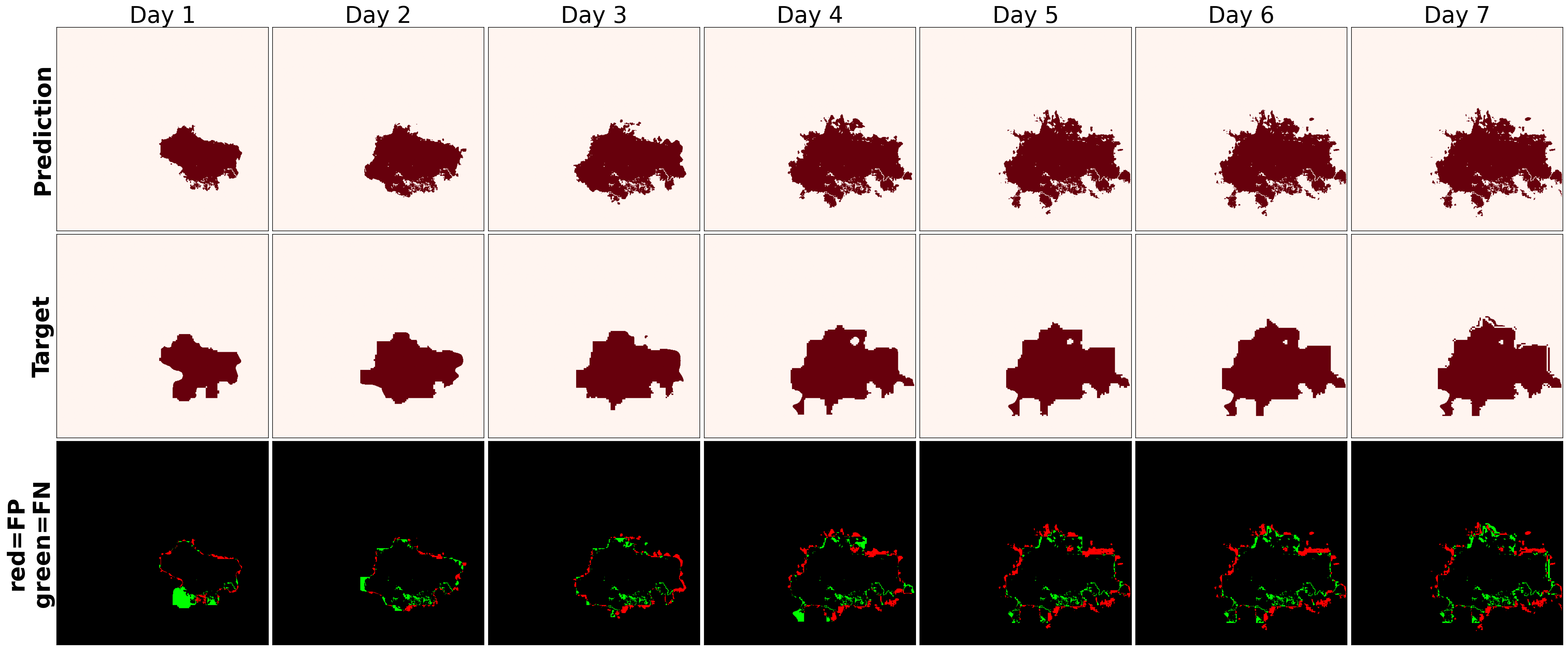}
	\caption{\revtwo{Buck 2017 fire analysis: (a) Days 1--7. Layout and color conventions as defined at the start of Section~\ref{subsec:spatial_analysis}.}}
	\label{fig:buck_combined}
\end{figure*}

\begin{figure*}[htbp]
	\ContinuedFloat
	\centering
	\includegraphics[width=0.9\textwidth]{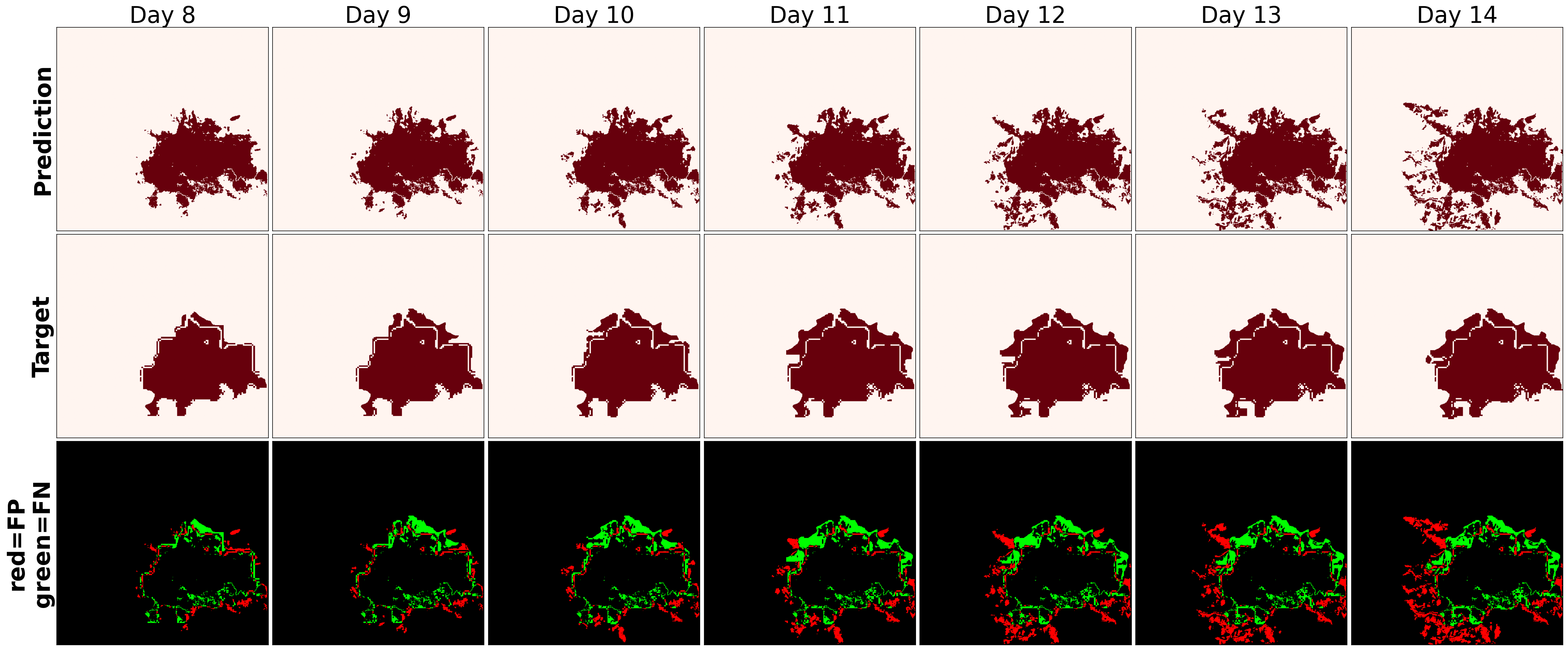}
	\caption{\revtwo{(cont.) Buck 2017. Days 8--14.}}
	\label{fig:buck_b}
\end{figure*}

\begin{figure*}[htbp]
	\ContinuedFloat
	\centering
	\includegraphics[width=0.9\textwidth]{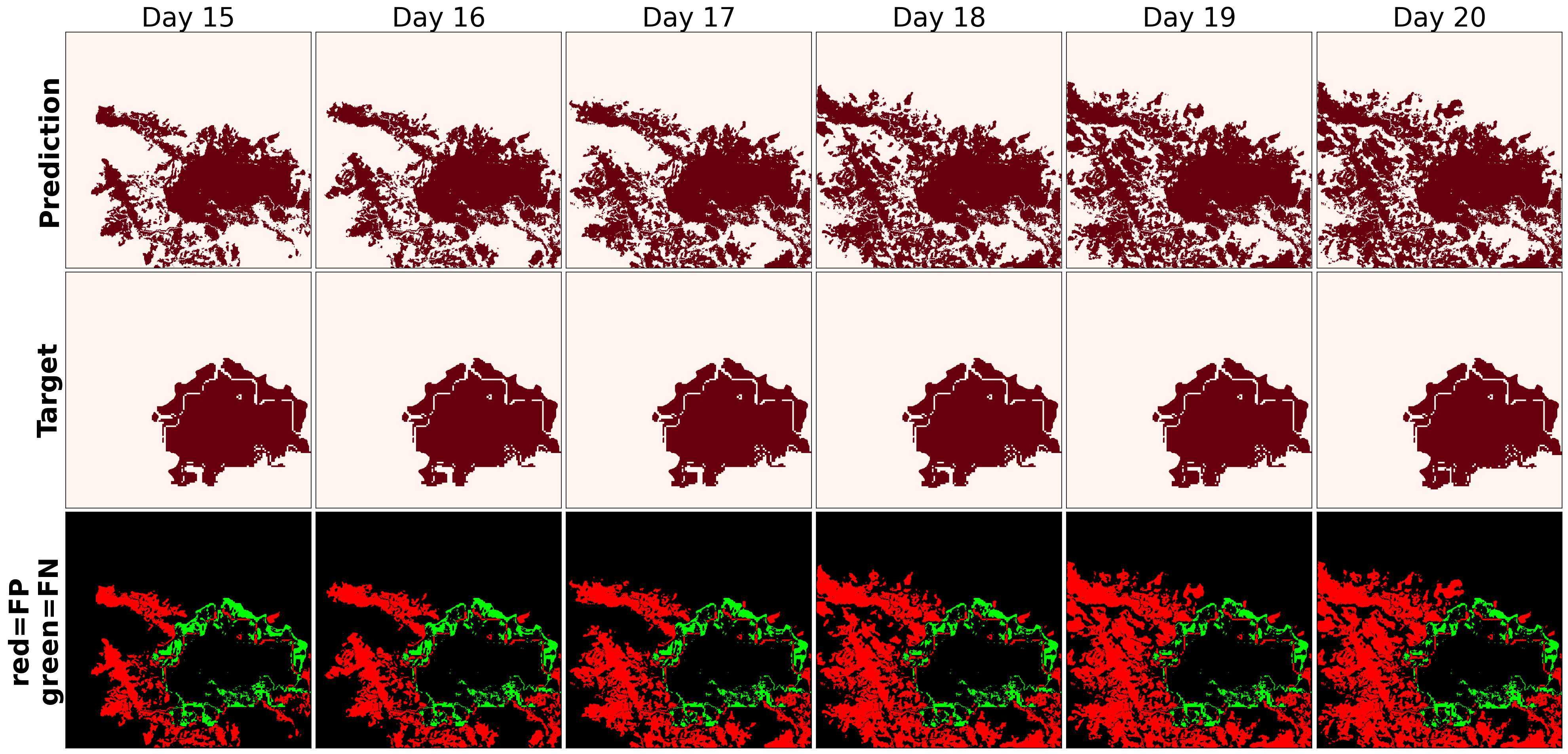}
	\caption{\revtwo{(cont.) Buck 2017. Days 15--20.}}
	\label{fig:buck_c}
\end{figure*}

\textbf{Buck 2017} shows (Figure \ref{fig:metrics}) a worse prediction accuracy \revtwo{(event-mean $IoU = 0.61$)} because the fire occurred within the Yolla Bolly - Middle Eel Wilderness \cite{calfire_buck2017}, a federally protected area where incident command applied Minimum Impact Suppression Tactics and passive containment without heavy machinery \cite{yubanet_buck2017}. The remaining five events were fought actively using direct or indirect attack strategies involving bulldozers, aviation, and large-scale firing operations. Since the model is trained on observed perimeters that inherently encode firefighting effects, it implicitly learns the suppression patterns dominant across the training data. The fundamentally different management regime of Buck 2017 produces perimeter dynamics that diverge from these learned patterns, explaining the degraded forecast accuracy. \revtwo{This management-regime mismatch is the dominant signal in Figure~\ref{fig:metrics}(c): Buck~2017 exhibits the steepest Precision decline of any event, falling from approximately 0.85 in the assimilation phase to roughly 0.35 by Day~20, while Recall remains stable around 0.75--0.85. The asymmetric degradation -- collapsing Precision but stable Recall -- is the quantitative signature of a simulated perimeter that increasingly outruns a passively contained observed perimeter, rather than of a model that loses sight of the fire altogether.}

\section{Conclusions}

We have introduced a hybrid approach for wildfire spread modeling that combines the physical interpretability of Probabilistic CA with the spatial feature extraction capacity of deep CNNs. By using an MS-CNN to generate spatially varying parameters, the model overcomes the homogeneous parameterization of traditional simulators. Crucially, the learned embeddings allow the system to accurately predict fire propagation well beyond standard vegetative boundaries, addressing the systemic failure point where up to 65.3\% of catastrophic fires burn outside of traditional canopy masks. Implemented in JAX, the framework exploits vectorized computation over large spatial domains. \revtwo{Several limitations of the present formulation should be made explicit. The CA contains no dedicated terms for ember spotting beyond the Moore neighborhood, for spontaneous secondary ignitions, or for active suppression. Because the network is trained against observed perimeters that are themselves shaped by these processes, the learned $fuel\_factor$ and, to a lesser extent, the other spatially varying parameters function as \emph{effective} quantities that conflate the propagation potential of the underlying fuel with the imprint of these unmodeled processes. Two of our event-level results illustrate this directly: the Ferguson~2018 forecast is degraded on early days because the secondary ignition cannot be reproduced by a strictly contagious CA (Figure~\ref{fig:ferguson_combined}), and the reduced Buck~2017 accuracy reflects a wilderness-area suppression regime that diverges from the active-suppression patterns implicitly learned from the other five events (Figure~\ref{fig:buck_combined}). Disentangling true fuel-driven propagation from these sub-grid contributions-for example, through an explicit spotting kernel, suppression covariates derived from incident records, or physics-informed regularization of the embedding-is an important direction for future work.} \revtwo{A second class of limitations concerns the scope of the evaluation. Our six events are concentrated in California and the immediately adjacent Lake County, Oregon, all sharing broadly similar Mediterranean and high-desert fuel and climate regimes, so transfer of the trained parameters to qualitatively different fire-prone regions (boreal, tropical, Australian eucalypt) is not demonstrated here. The methodology itself is geographically agnostic-the static (LANDFIRE) and dynamic (ERA5) input modalities are available across the conterminous United States and globally, respectively-but an empirical cross-regional transfer study is left to future work. A related limitation is that our quantitative benchmarks are concentrated on the two prior differentiable-CA frameworks \cite{xia2025pytorchfire, cakir2025jaxwildfire}, which share our task formulation, input modalities, and evaluation dataset \cite{xia2025pytorchfire_data} and so admit a controlled head-to-head comparison on the same events. Recently proposed data-driven spread approaches \cite{lahrichi2025improved, anastasiou2025wildfire, yu2025probabilistic, zhou2025physfire, xu2026advanced} address adjacent but distinct problem formulations (coarse-grid pixel classification, denoising-diffusion samples, graph neural ODEs over global fire activity) and were therefore positioned in the introduction rather than benchmarked numerically; a controlled multi-paradigm benchmark on perimeter-trajectory data is a clear next step.} Future work includes integration into operational disaster management systems and reinforcement learning environments for autonomous fire suppression.

\section*{Data availability}
We provide the code and data used in this study at the following link: \url{https://github.com/mzhen77/neural-ca-wildfire}.

\bibliographystyle{cas-model2-names}
\bibliography{ref}

@techreport{rothermel1972mathematical,
	author      = {Rothermel, Richard C.},
	title       = {A mathematical model for predicting fire spread in wildland fuels},
	year        = {1972},
	institution = {USDA Forest Service, Intermountain Forest and Range Experiment Station},
	type        = {Research Paper},
	number      = {INT-115}
}

@techreport{finney1998farsite,
	author      = {Finney, Mark A.},
	title       = {{FARSITE}: Fire Area Simulator-Model Development and Evaluation},
	year        = {1998},
	institution = {USDA Forest Service, Rocky Mountain Research Station},
	type        = {Research Paper},
	number      = {RMRS-RP-4},
	address     = {Ogden, UT}
}

@inproceedings{finney2006flammap,
	author    = {Finney, Mark A.},
	title     = {An overview of {FlamMap} fire modeling capabilities},
	booktitle = {Fuels Management-How to Measure Success: Conference Proceedings},
	year      = {2006},
	editor    = {Andrews, Patricia L. and Butler, Bret W.},
	series    = {RMRS-P-41},
	pages     = {213--220},
	publisher = {USDA Forest Service, Rocky Mountain Research Station},
	address   = {Portland, OR}
}

@article{finney2002mtt,
	author  = {Finney, Mark A.},
	title   = {Fire growth using minimum travel time methods},
	journal = {Canadian Journal of Forest Research},
	volume  = {32},
	number  = {8},
	pages   = {1420--1424},
	year    = {2002},
	doi     = {10.1139/x02-068}
}

@article{alexandridis2008cellular,
	author  = {Alexandridis, A. and Vakalis, D. and Siettos, C. I. and Bafas, G. V.},
	title   = {A cellular automata model for forest fire spread prediction: The case of the wildfire that swept through {Spetses Island} in 1990},
	journal = {Applied Mathematics and Computation},
	volume  = {204},
	number  = {1},
	pages   = {191--201},
	year    = {2008},
	doi     = {10.1016/j.amc.2008.06.046}
}

@article{freire2019cellular,
	author  = {Freire, Jos{\'e} G. and DaCamara, Carlos C.},
	title   = {Using cellular automata to simulate wildfire propagation and to assist in fire management},
	journal = {Natural Hazards and Earth System Sciences},
	volume  = {19},
	number  = {1},
	pages   = {169--179},
	year    = {2019},
	doi     = {10.5194/nhess-19-169-2019}
}

@article{zheng2017forest,
	author  = {Zheng, Zhong and Huang, Wei and Li, Songnian and Zeng, Yongnian},
	title   = {Forest fire spread simulating model using cellular automaton with extreme learning machine},
	journal = {Ecological Modelling},
	volume  = {348},
	pages   = {33--43},
	year    = {2017},
	doi     = {10.1016/j.ecolmodel.2016.12.022}
}

@article{xia2025pytorchfire,
	author  = {Xia, Zeyu and Cheng, Sibo},
	title   = {{PyTorchFire}: A {GPU}-accelerated wildfire simulator with differentiable cellular automata},
	journal = {Environmental Modelling \& Software},
	volume  = {188},
	pages   = {106401},
	year    = {2025},
	issn    = {1364-8152},
	doi     = {10.1016/j.envsoft.2025.106401}
}

@misc{xia2025pytorchfire_data,
	author       = {Xia, Zeyu and Cheng, Sibo},
	title        = {Data for: {PyTorchFire}: A {GPU}-accelerated wildfire simulator with differentiable cellular automata},
	year         = {2024},
	howpublished = {Mendeley Data, V1},
	doi          = {10.17632/nx2wsksp9k.1},
	note         = {Dataset}
}

@article{cakir2025jaxwildfire,
	author  = {{\c{C}}ak{\i}r, Ufuk and Darvariu, Victor-Alexandru and Lacerda, Bruno and Hawes, Nick},
	title   = {{JaxWildfire}: A {GPU}-accelerated wildfire simulator for reinforcement learning},
	journal = {arXiv preprint arXiv:2512.06102},
	year    = {2025},
	url     = {https://arxiv.org/abs/2512.06102}
}

@article{weinhouse2025leveraging,
	author  = {Weinhouse, Connor and Augustin, J.},
	title   = {Leveraging cellular automata for real-time wildfire spread modeling in {California}},
	journal = {arXiv preprint arXiv:2510.09708},
	year    = {2025},
	url     = {https://arxiv.org/abs/2510.09708}
}

@article{jain2020review,
	author  = {Jain, Piyush and Coogan, Sean C. P. and Subramanian, Sriram G. and Crowley, Mark and Taylor, Steve and Flannigan, Mike D.},
	title   = {A review of machine learning applications in wildfire science and management},
	journal = {Environmental Reviews},
	volume  = {28},
	number  = {4},
	pages   = {478--505},
	year    = {2020},
	doi     = {10.1139/er-2020-0019}
}

@article{shen2023differentiable,
	author  = {Shen, Chaopeng and Appling, Alison P. and Gentine, Pierre and Bandai, Toshiyuki and Gupta, Hoshin and Tartakovsky, Alexandre and others},
	title   = {Differentiable modelling to unify machine learning and physical models for geosciences},
	journal = {Nature Reviews Earth \& Environment},
	volume  = {4},
	number  = {8},
	pages   = {552--567},
	year    = {2023},
	doi     = {10.1038/s43017-023-00450-9}
}

@article{chen2024autost,
	author  = {Chen, Xuexue and Tian, Ye and Zheng, Change and Liu, Xiaodong},
	title   = {{AutoST-Net}: A spatiotemporal feature-driven approach for accurate forest fire spread prediction from remote sensing data},
	journal = {Forests},
	volume  = {15},
	number  = {4},
	pages   = {705},
	year    = {2024},
	doi     = {10.3390/f15040705}
}

@article{lahrichi2025improved,
	author  = {Lahrichi, Saad and Bova, Jake and Johnson, Jesse and Malof, Jordan},
	title   = {Improved wildfire spread prediction with time-series data and the {WSTS+} benchmark},
	journal = {arXiv preprint arXiv:2502.12003},
	year    = {2025},
	url     = {https://arxiv.org/abs/2502.12003}
}

@article{anastasiou2025wildfire,
	author  = {Anastasiou, Nikolaos and Kondylatos, Spyros and Papoutsis, Ioannis},
	title   = {Wildfire spread forecasting with deep learning},
	journal = {arXiv preprint arXiv:2505.17556},
	year    = {2025},
	url     = {https://arxiv.org/abs/2505.17556}
}

@article{yu2025probabilistic,
	author  = {Yu, Wenbo and Ghosh, Anirbit and Finn, Tobias Sebastian and Arcucci, Rossella and Bocquet, Marc and Cheng, Sibo},
	title   = {A probabilistic approach to wildfire spread prediction using a denoising diffusion surrogate model},
	journal = {arXiv preprint arXiv:2507.00761},
	year    = {2025},
	url     = {https://arxiv.org/abs/2507.00761}
}

@article{zhou2025physfire,
	author  = {Zhou, Nan and Wang, Huandong and Li, Jiahao and Li, Yang and Zhang, Xiao-Ping and Li, Yong and Chen, Xinlei},
	title   = {{PhysFire-WM}: A physics-informed world model for emulating fire spread dynamics},
	journal = {arXiv preprint arXiv:2512.17152},
	year    = {2025},
	url     = {https://arxiv.org/abs/2512.17152}
}

@article{xu2026advanced,
	author  = {Xu, Fan and Gong, Wei and Wu, Hao and Peng, Lilan and Zhao, Xibin and others},
	title   = {Advanced global wildfire activity modeling with hierarchical graph {ODE}},
	journal = {arXiv preprint arXiv:2601.01501},
	year    = {2026},
	url     = {https://arxiv.org/abs/2601.01501}
}

@article{mathur2026spatiotemporal,
	author  = {Mathur, Shaurya and Manjunath, Shreyas Bellary and Kulkarni, Nitin and Vereshchaka, Alina},
	title   = {Spatiotemporal wildfire prediction and reinforcement learning for helitack suppression},
	journal = {arXiv preprint arXiv:2601.14238},
	year    = {2026},
	url     = {https://arxiv.org/abs/2601.14238}
}

@misc{jax2018github,
	author  = {Bradbury, James and Frostig, Roy and Hawkins, Peter and Johnson, Matthew James and Leary, Chris and Maclaurin, Dougal and Necula, George and Paszke, Adam and VanderPlas, Jake and Wanderman-Milne, Skye and Zhang, Qiao},
	title   = {{JAX}: Composable transformations of {Python+NumPy} programs},
	year    = {2018},
	version = {0.9.1},
	url     = {http://github.com/jax-ml/jax}
}

@misc{joaquin2019era5land,
	author       = {Mu{\~n}oz Sabater, Joaqu{\'i}n},
	title        = {{ERA5-Land} monthly averaged data from 1981 to present},
	year         = {2019},
	howpublished = {Copernicus Climate Change Service (C3S) Climate Data Store (CDS)},
	doi          = {10.24381/cds.68d2bb30}
}

@misc{calfire_2020_incidents,
	author       = {{California Department of Forestry and Fire Protection}},
	title        = {2020 Fire Season Incident Archive},
	year         = {2020},
	howpublished = {\url{https://www.fire.ca.gov/incidents/2020}}
}

@misc{calfire_buck2017,
	author       = {{California Department of Forestry and Fire Protection}},
	title        = {Buck Fire Incident Report},
	year         = {2017},
	howpublished = {\url{https://www.fire.ca.gov/incidents/2017/9/12/buck-fire}}
}

@misc{yubanet_buck2017,
	author       = {{Shasta Trinity National Forest}},
	title        = {Buck Fire - Incident Status Updates},
	year         = {2017},
	howpublished = {\url{https://yubanet.com/containedca16/buck/}}
}

\end{document}